\documentclass[a4paper,12pt,number]{elsarticle}

\usepackage{amssymb,amsmath}
\usepackage{graphicx}
\usepackage{geometry}
\usepackage{xcolor}
\usepackage[normalsize]{subfigure}
\usepackage{array}
\usepackage{ulem}
\usepackage{hyperref}

\newcommand{\Eref}[1]{Eq.~\eqref{#1}}

\newcommand{\Fref}[1]{Fig.~\ref{#1}}
\newcommand{\Tref}[1]{Tab.~\ref{#1}}
\newcommand{\ASref}[1]{{\bf Appendix~\ref{#1}}}

\newcommand{\save}[1]{}

\newcommand{\rev}[2]{#1}

\newcommand{\code}[1]{{\bf #1}}

\newcommand{\Sref}[1]{{\bf Section~\ref{#1}}}

\newcommand{\http}[1]{{\tt http://#1}}

\newcommand\de[1]{\,{\mathrm d}#1}
\newcommand\vek[1]{\mathbf{#1}} 

\newcommand{\bmath}[1]{\mbox{\boldmath$#1$}}
\newcommand{\tenss}[1]{\bmath{#1}} 
\newcommand{\tensf}[1]{\bmath{\mathbf{#1}}} 
\newcommand{\tensd}[1]{\bmath{\mathcal{#1}}} 

\newcommand{\trn}{{\sf T}}

\newcommand{\norm}[1]{\|#1\|}

\newcommand{\semtrx}[1]{\mathsf{#1}} 
\newcommand{\sevek}[1]{\mathsf{#1}} 

\newcommand{\dcontr}{\,\colon}

\newcommand{\strain}{\varepsilon}
\newcommand{\vstrain}{\epsilon} 
\newcommand{\vstress}{\sigma}   
\newcommand{\stress}{\sigma}

\newcommand\abs[1]{|#1|} 

\newcommand{\IncDomain}{\Omega}

\newcommand{\myemph}[1]{\emph{#1}}
\newcommand{\aname}[1]{#1} 
\newcommand{\mname}[1]{\emph{#1}} 

\newcounter{alglinenumber}
\newcommand{\algstep}{\hspace{5mm}}
\newcommand{\algnum}{\addtocounter{alglinenumber}{1}{\scriptsize \sf \thealglinenumber}}
\newcommand{\algline}[1]{\algnum& #1 \\} 
\newcommand{\function}[1]{{\bf #1}\,}

\newenvironment{algorithm}[1]{%
\setcounter{alglinenumber}{0}
\begin{center}
\begin{tabular}{|rl|}
\hline
& \function{#1} \\
}{ 
\hline
\end{tabular}
\end{center}
\vspace*{-4mm}
}

\newcommand{\lline}[1]{{\bf Line #1}}

\newcommand{\x}{\vek{x}}
\newcommand{\half}{\mbox{$\frac{1}{2}$}} 

\journal{Computer Methods in Applied Mechanics and Engineering}
\begin{document}

\begin{frontmatter}

\title{A micromechanics-enhanced finite element formulation for modelling heterogeneous materials}

\author[gu,ctu,unsw]{Jan Nov\'{a}k}
\ead{novakj@cml.fsv.cvut.cz}
\cortext[cor]{Corresponding author. Tel.:~+44-141-330-5207}

\author[gu]{{\L}ukasz Kaczmarczyk}
\ead{Lukasz.Kaczmarczyk@glasgow.ac.uk}
\author[gu]{Peter Grassl}
\ead{Peter.Grassl@glasgow.ac.uk}
\author[ctu]{Jan Zeman}
\ead{zemanj@cml.fsv.cvut.cz}
\author[gu]{Chris J. Pearce\corref{cor}}
\ead{Chris.Pearce@glasgow.ac.uk}

\address[gu]{School of Engineering, University of Glasgow, Glasgow G12 8QQ, UK}
\address[ctu]{Faculty of Civil Engineering, Czech Technical University
  in Prague, Th\'{a}kurova 7, \mbox{166 29 Praha 6}, Czech Republic}
\address[unsw]{School of Civil and Environmental Engineering, University of New
South Wales, NSW~2052, Sydney, Australia}

\begin{abstract}
In the analysis of composite materials with heterogeneous microstructures, full
resolution of the heterogeneities using classical numerical approaches can be
computationally prohibitive. This paper presents a micromechanics-enhanced
finite element formulation that accurately captures the mechanical behaviour of
heterogeneous materials in a computationally efficient manner. The strategy
exploits analytical solutions derived by Eshelby for
ellipsoidal inclusions in order to determine the mechanical perturbation fields
as a result of the underlying heterogeneities. Approximation functions for these
perturbation fields are then incorporated into a finite element formulation to
augment those of the macroscopic fields. A significant feature of this approach
is that the finite element mesh does not explicitly resolve the heterogeneities
and that no additional degrees of freedom are introduced. In this paper,
hybrid-Trefftz stress finite elements are utilised and performance of the
proposed formulation is demonstrated with numerical examples. The method is
restricted here to elastic particulate composites with ellipsoidal inclusions
but it has been designed to be extensible to a wider class of materials
comprising arbitrary shaped inclusions.
\end{abstract}


\end{frontmatter}

\section{Introduction}
%
In the analysis of materials with complex microstructures, full resolution of
the heterogeneities using classical numerical approaches such as the Finite
Element method can be computationally prohibitive. To overcome this, one option
is to model the macroscale problem using equivalent properties; however, this
can lead to a critical loss of information about the finer scale behaviour and
poor understanding of the heterogeneities' influence on the macroscale response.
Numerical approaches such as computational homogenization (often called
$\textrm{FE}^2$) provide an alternative strategy~\cite{Feyel2000309,Geers2010}.
These techniques comprise nested Finite Element analyses, where each macroscopic
material point response is determined via the numerical solution of an
RVE subject to the macroscopic strains. Although such approaches have
significant potential for certain classes of problems, they are still
computationally demanding and are restricted to situations involving clear
separation of scales.

The objective of this work is to develop a Finite Element formulation for
modelling the macroscopic mechanical problem that is enhanced to capture the
influence of the underlying heterogeneities. In our approach, the Finite Element
mesh is not required to explicitly resolve the heterogeneities. Closed-form
expressions for the perturbation of the mechanical fields due to the presence of
the heterogeneities are determined and these are then utilised to enhance the
Finite Element formulation.

The ability to capture the effect of microstructural features
independently of the underlying finite element mesh has been an ongoing
challenge in computational mechanics research. Partition of Unity
methods~\cite{Sukumar:2001:MHI,Moes:1999:FEM,Moes:2003:CAHC,Wells2002} provide a potential
solution to this problem, without mesh refinement, by extending a given solution
space with additional functions and has been successfully applied to problems
such as cracks and material interfaces. The application of this approach in the
context of the current work will be briefly discussed in this paper, whereby the
closed-form solutions derived for the mechanical perturbation fields are used to
extend the classical finite element method. However, it will be shown that there
are some disadvantages to this approach for the particular problem at hand and
an alternative approach, centred on the Hybrid-Trefftz stress element
formulation~\cite{Kaczmarczyk20091298}, represents the main focus of this paper.
This method does not result in additional degrees of freedom, although it does
involve an additional, albeit relatively minor, computational overhead.

The heterogeneities, although currently restricted to simple shapes
(ellipsoids), can be randomly sized and randomly distributed without reference
to the finite element mesh. Therefore, the proposed approach has the potential
to be applicable to a wide range of composite materials, such as fibre
reinforced composites~\cite{Kabele2007194}, porous
media~\cite{nicolaou2004hybrid,Fritsch2009230}, functionally graded
materials~\cite{sharif2008microstructure}, etc. Moreover, it can be extended to
general inclusion shapes by evaluating the perturbation functions
numerically~\cite{maz2007approximate,Novak:2008:CESS}.

The paper is structured as follows. The methodology of the proposed strategy is
described in \Sref{s:methodology}. Construction of the perturbation
approximation functions for Finite Element Analysis is derived in
\Sref{FEA_perturbation}. The implementation into the Hybrid-Trefftz stress
element formulation containing an arbitrary number of inclusions is presented in
\Sref{s:hts-formulation}. \Sref{s:results} comprises examples demonstrating the
model's performance. Finally we present the conclusions as well as a discussion
on future research directions. An appendix is included that highlights some
important, but rather technical, aspects of the proposed technique in order to
keep the paper self-contained.

\section{Micromechanics approach}\label{s:methodology}
%
This section outlines the strategy to calculate the perturbation of mechanical
fields due to a heterogeneous microstructure, exploiting the \mname{Equivalent
Inclusion Method}~\cite{mura1987micromechanics} in conjunction with analytical
micromechanics. Our objective is to convert the heterogeneous problem into an
equivalent homogeneous problem and to derive analytical expressions for the
perturbations of the stress, strain and displacement fields that we can then
utilise within a finite element formulation.

\begin{figure}[!ht]
  \centering
    \begin{tabular}{>{\centering}m{50mm}>{\centering}m{0mm}>{\centering\arraybackslash}m{50mm}}
      \includegraphics[width=50mm]{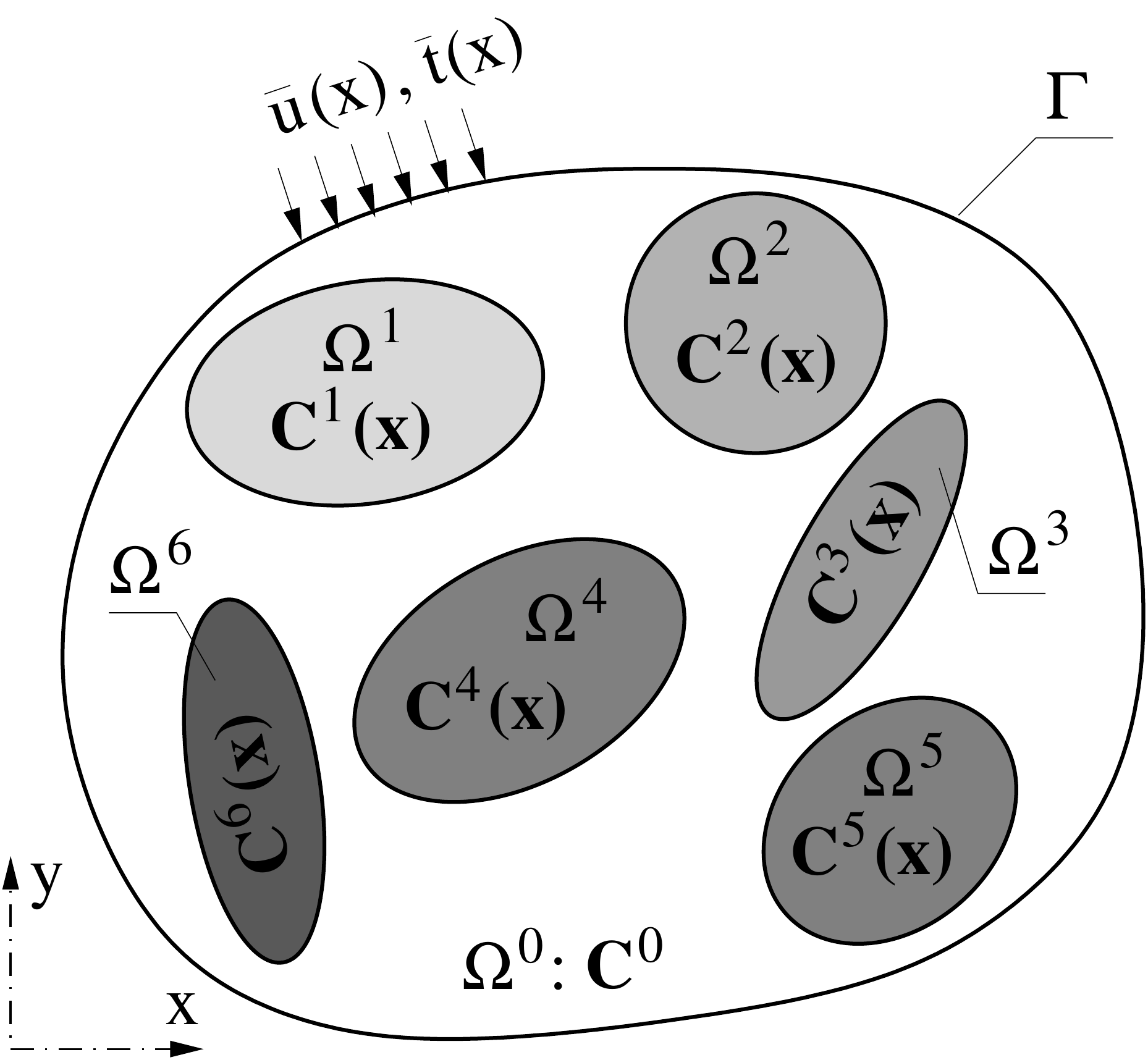} &$\approx$
      &\includegraphics[width=50mm]{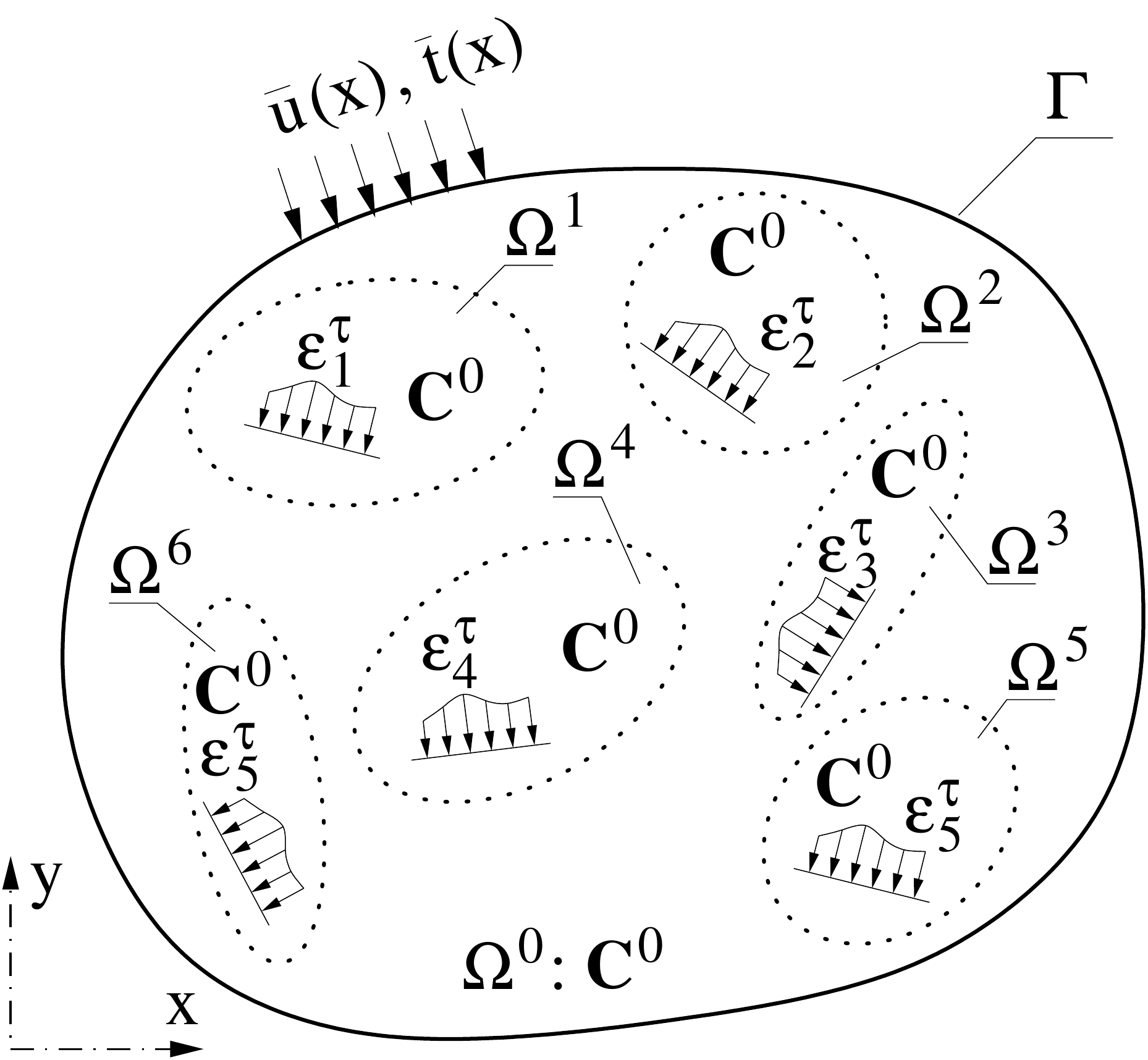}\\
      (a) & &(b)
    \end{tabular}
    \caption{Principle of \mname{Equivalent Inclusion Method}: a)
      composite body with inclusions, b) homogeneous reference body
      with additional equivalent eigenstrains}
    \label{fig:approx_fces_solu_strategy}
\end{figure}

Consider a body consisting of clearly distinguishable heterogeneities in a
matrix (\Fref{fig:approx_fces_solu_strategy}a) subjected to a displacement
$\vek{\overline{u}}$ and traction $\vek{\overline{t}}$ field.
The stiffness of such a
material is
decomposed as follows~\cite{mura1987micromechanics,Novak:2008:CESS}
\begin{equation}
  \tensf{C}
  =
  \tensf{C}^0
  +
  V \tensf{C}^*
  \label{eq:stiff_decomposition}
\end{equation}
where $\tensf{C}^0 $ is the stiffness tensor of a homogeneous matrix $\Omega^0$
and $\tensf{C}^* = \sum^N_i [\tensf{C}^i - \tensf{C}^0]$ is due to the presence
of $N$ inclusions. $\tensf{C}^*$ is nonzero only within the domain $\Omega =
\Omega^1\cup\dots \cup\Omega^N$, so that
\begin{equation}
  V
  =
  \left\{
  \begin{array}{ll}
    0 \mbox{ in } \Omega^0\\
    1 \mbox{ in } \Omega
  \end{array}\right.
\end{equation}
As a result of the heterogeneities, the mechanical fields (displacement, strains, stresses) experience a perturbation for which we will derive closed-form expressions based on analytical micromechanics. Symbolically, we can express the decomposition of the mechanical fields as follows:
\begin{eqnarray}
  \vek{u}
  =
  \vek{u}^0
  +
  \vek{u}^*,
  &
  \tenss{\strain}
  =
  \tenss{\strain}^0
  +
  \tenss{\strain}^*,
  &
  \tenss{\stress}
  =
  \tenss{\stress}^0
  +
  \tenss{\stress}^*
  \label{eq:fields_decomposition}
 \end{eqnarray}
where, the superscript $\cdot^0$ indicates the \myemph{macroscopic} component of
the fields in the absence of heterogeneities and superscript $\cdot^*$ indicates
the \myemph{perturbation} (or \myemph{microscopic}) component due to the
presence of the heterogeneities. It is worth noting that, traditionally, in
analytical micromechanics, the \myemph{macroscopic} fields are assumed to be
uniform across the domain, e.g.~\cite{Vorel:2009:SEM,pichler2008micron}.
Here it is assumed that they can be position dependent functions of the
\aname{Neumann} and \aname{Dirichlet} boundary conditions.

The perturbation fields are determined by employing the \mname{equivalent
inclusion method} for a single heterogeneity embedded in a matrix and then
extended here for multiple heterogeneities. In the \mname{equivalent inclusion
method}, the heterogeneous solid is replaced by an equivalent homogeneous solid
with uniform material stiffness $\tensf{C}^0$ everywhere
(\Fref{fig:approx_fces_solu_strategy}a, b) and suitable stress-free eigenstrains
$\tenss{\strain}^{\tau}_i$ applied in the inclusions $\Omega^i$ so that
the homogeneous equivalent solid has the same mechanical fields as the original
heterogeneous solid.

\subsection{Equivalent inclusion method for single heterogeneity problem}
\begin{figure}[ht]
  \centering
    \begin{tabular}{>{\centering}m{40mm}>{\centering}m{0mm}>{\centering}m{40mm}>{\centering}m{0mm}%
        >{\centering\arraybackslash}m{40mm}}
       \includegraphics[width=45mm]{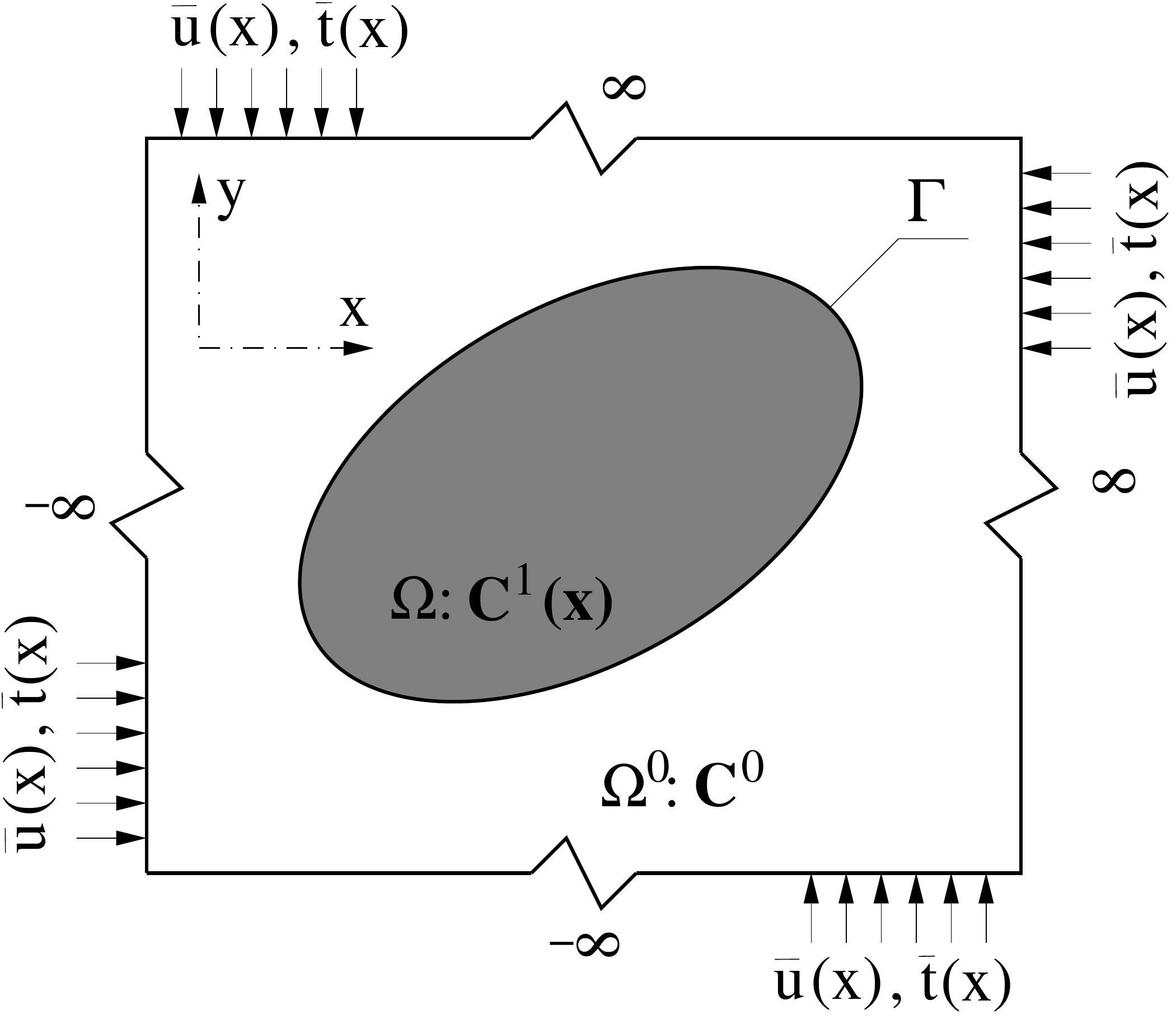} &$\equiv$
       &\includegraphics[width=45mm]{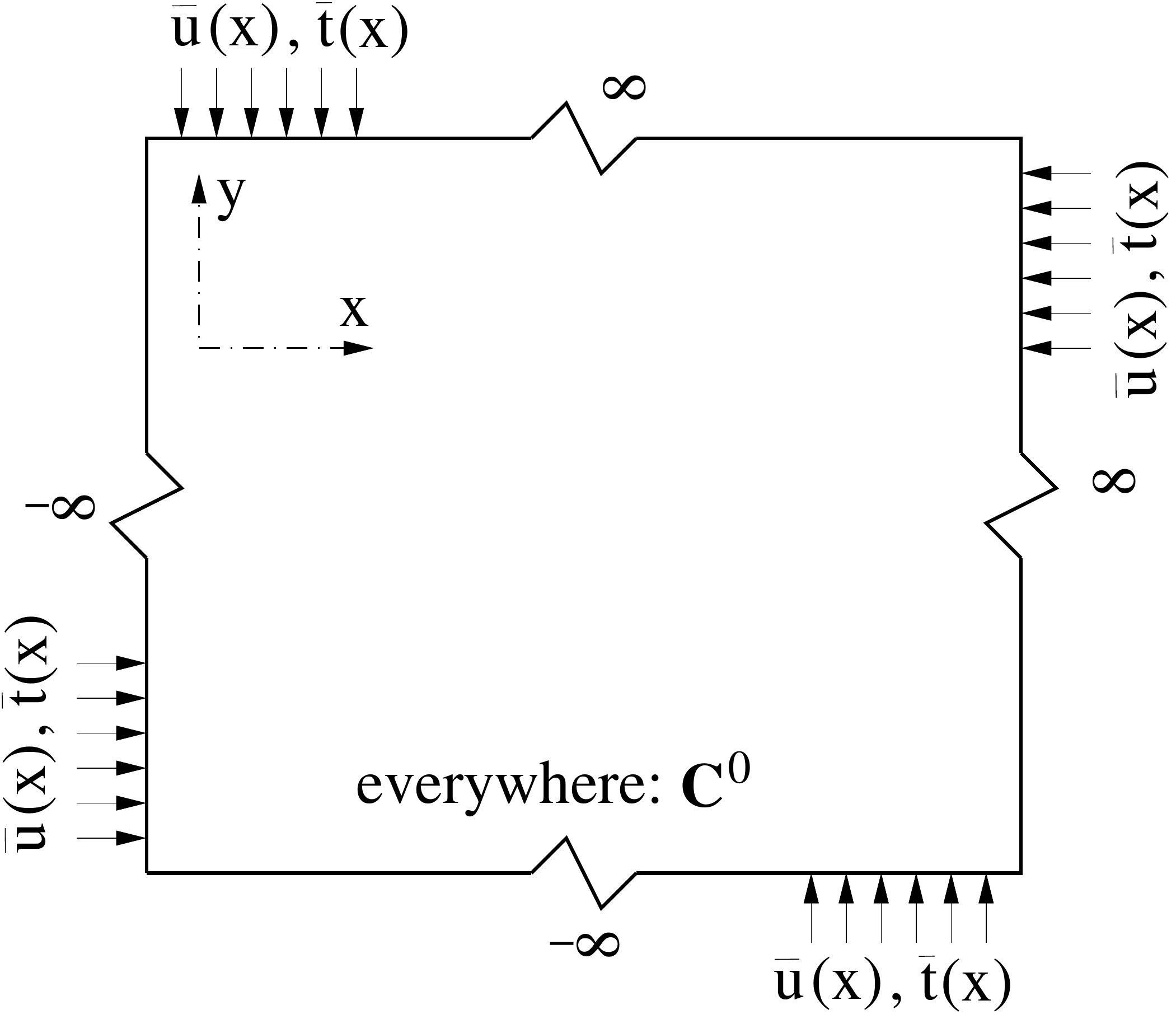}
       &$+$ &\includegraphics[width=45mm]{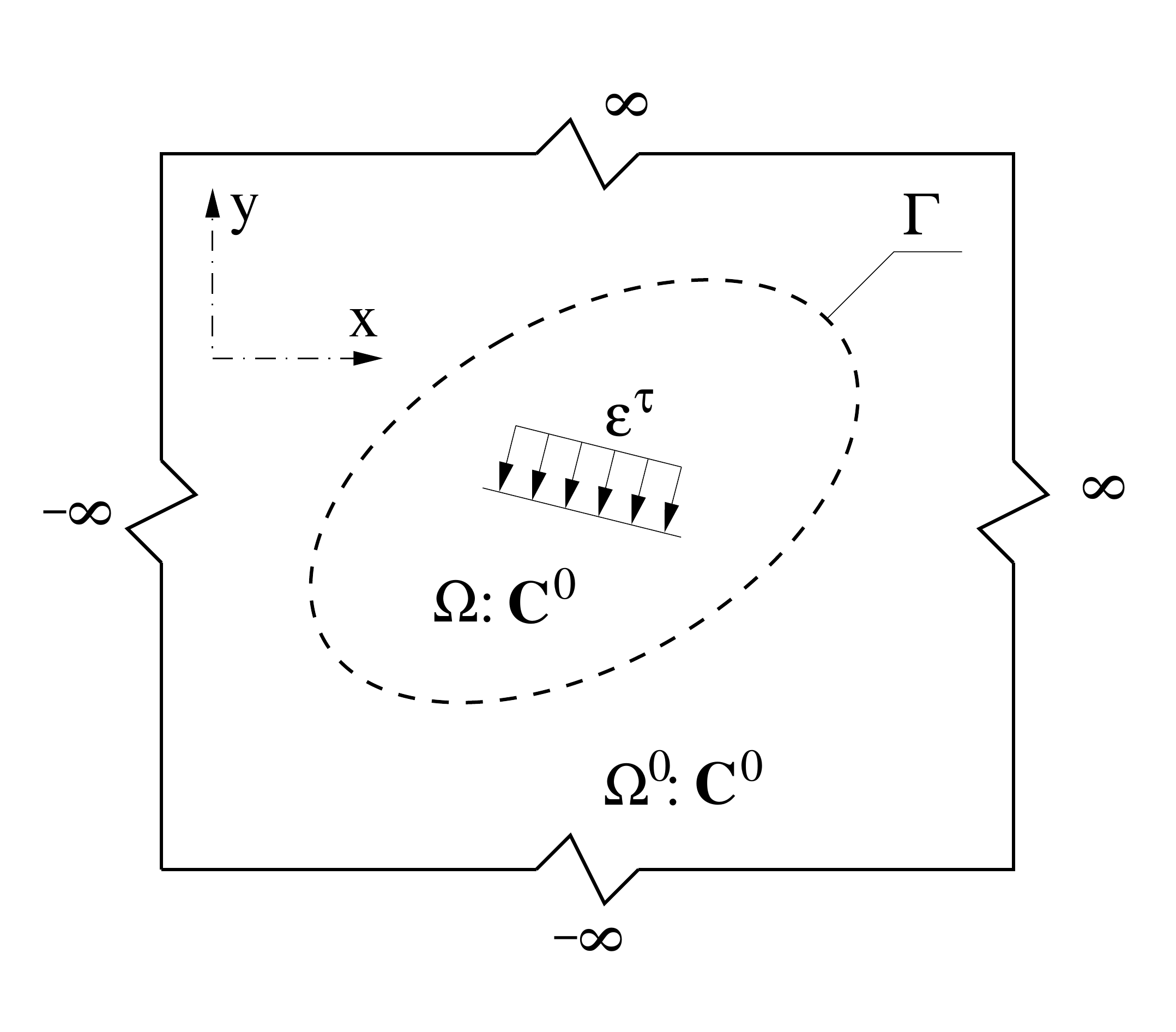}\\
      (a) & &(b) & &(c)
    \end{tabular}
    \caption{\mname{Equivalent Inclusion Method}: a)
      inhomogeneity problem, b) infinite homogeneous body,
      c) homogeneous inclusion problem}
    \label{figA:eq_incl_method_principle}
\end{figure}

Consider first a single heterogeneity embedded in a matrix. Following \aname{Eshelby's}
fundamental work~\cite{eshelby1957determination},
this problem can be decomposed into two problems of known solution and then
assembled back via superposition~\cite{mura1987micromechanics,eshelby1957determination}, see~\Fref{figA:eq_incl_method_principle}.
In brief, the solution of the inhomogeneity problem requires the determination
of the transformation eigenstrain
$\tenss{\strain}^{\tau}$ that induces the identical local
mechanical response as the original heterogeneous body. \rev{Due to the absence of other inclusions, no strain or stress concentrations are induced and thus $\tenss{\strain}^{\tau}$ remains constant~\cite{eshelby1957determination}}{}.

In the original heterogeneous body, the stress state can be expressed as
\begin{equation}
\label{stress_het}
  \tenss{\stress}
  =
  \tenss{\stress}^0
  +
  \tenss{\stress}^*
  =
  \tensf{C}
  \dcontr[
    \tenss{\strain}^0
    +
    \tenss{\strain}^*
    ]
\end{equation}
In the homogeneous solid, we add a stress-free eigenstrain
$\tenss{\strain}^\tau$ inside the domain of the inclusion, which has the same
material stiffness $\tensf{C}^0$ as the matrix such that
\begin{equation}\label{stress_hom}
\tenss{\stress}
=
\tensf{C}^0
\dcontr
[\tenss{\strain}^0
+
\tenss{\strain}^*
-
\tenss{\strain}^\tau]
\end{equation}
It should be noted that $\tenss{\strain}^\tau=0$ in the matrix.

Given that the macroscopic stress is $\tenss{\stress}^0 =
\tensf{C}^0\dcontr \tenss{\strain}^0$, it can be see from equating
Eqs.~(\ref{stress_het}) and (\ref{stress_hom}), that the stress perturbation in
the homogeneous solid can be expressed as
\begin{equation}\label{stress_per}
\tenss{\stress}^*
=
\tensf{C}^0
\dcontr
[\tenss{\strain}^*
-
\tenss{\strain}^\tau]
\end{equation}
Furthermore, equating Eqs.~(\ref{stress_het}) and (\ref{stress_hom}) also
results in the following expression:
\begin{equation}\label{stress_equality}
\tensf{C}
\dcontr
[\tenss{\strain}^0
+
\tenss{\strain}^*]
=
\tensf{C}^0
\dcontr
[\tenss{\strain}^0
+
\tenss{\strain}^*
-
\tenss{\strain}^\tau]
\end{equation}
where the transformation eigenstrain is as yet unknown. Eshelby's
solution of the homogeneous inclusion problem~\cite{eshelby1957determination},
relates the eigenstrain to the perturbation strain as follows
\begin{equation}\label{eshelby}
\tenss{\strain}^*
=
\tensf{S}
\dcontr
\tenss{\strain}^\tau
\end{equation}
where $\tensf{S}$ denotes the Eshelby tensor and is a function of the
heterogeneity's geometry and the material stiffness of both the matrix and
heterogeneity. Substituting this expression into Eq.~(\ref{stress_equality}) and
rearranging yields
\begin{equation}\label{eqA:eq_incl_meth_step_4}
\left[
  \tensf{C}
  -
  \tensf{C}^0
\right]
\dcontr
\tenss{\strain}^0
=
\left[
  \tensf{C}^0
  \dcontr
  \tensf{S}
  -
  \tensf{C}
  \dcontr
  \tensf{S}
  -
  \tensf{C}^0
\right]
\dcontr
\tenss{\strain}^\tau
\end{equation}
This can be recast in a compact form to give an expression for the eigenstrain
$\tenss{\strain}^\tau$ which depends on the homogeneous strain
$\tenss{\strain}^0$, the material stiffness of both the matrix
and heterogeneity and the Eshelby tensor as
\begin{equation}\label{eqA:eq_incl_meth_step_5}
\tenss{\strain}^\tau
=
\tensf{B}
\dcontr
\tenss{\strain}^0
\end{equation}
where the tensor $\tensf{B}$ is provided by:
\begin{equation}\label{eqA:remote_transf_strain_oper}
\tensf{B}
=
-\left[
  \tensf{C}^*
  \dcontr
  \tensf{S}
  +
  \tensf{C}^0
\right]^{-1}
\dcontr
\tensf{C}^*
\end{equation}
Once the transformation eigenstrain has been determined, the stress
perturbation can be computed from  Eq.~(\ref{stress_per}) in the form
\begin{equation}\label{stress_per_2}
\tenss{\stress}^*
=
\tensf{C}^0
\dcontr
[\tensf{S}
-
\tensf{I}]
\dcontr
\tensf{B}
\dcontr
\tenss{\strain}^0
\end{equation}
It can be seen that the stress perturbation depends on stiffness of the
different material phases, the macroscopic strain field and the geometry of the
heterogeneity. This closed-form expression for the stress perturbation is at the
heart of the proposed finite element enrichment to be discussed later in this
paper. It is also useful to derive an expression for the displacement
perturbation field as follows
\begin{equation}
\vek{u}^*
=
\tensd{L}
\dcontr
\tenss{\strain}^\tau
=
\tensd{L}
\dcontr
\tensf{B}
\dcontr
\tenss{\strain}^0
\end{equation}
where the operator $\tensd{L}$ is a third order tensor,
mapping $\tenss{\strain}^\tau \to \vek{u}^*$. For the sake of
conciseness and to keep the paper self-contained, the detailed derivation of
this operator can be found in~\ASref{appendix:B}.

\subsection{Self-compatibility algorithm for multiple inclusions}
%
In the case of multiple inclusions, the mechanical perturbation fields within individual inclusion domains are \rev{no longer uniformly distributed as a result of their mutual interaction. Here, we account for this approximately by assuming that the eigenfields to be piecewice constant within each inclusion. Thus, the perturbations are}{essentially} determined from the \aname{Eshelby} solution for each individual inclusion, as described above, \rev{plus}{However, it is also necessary to introduce} an iterative \rev{\myemph{self-compatibility}}{\myemph{self-balancing}} procedure (\Tref{alg:selfBalancing}) to ensure that the solution correctly reflects the influence of multiple \rev{heterogeneities}{inclusions}.

This procedure iteratively \rev{enforces compatibility (see eg.~\cite{pichler2010estimation} for further reference) between the imposed macroscopic strain and the average}{modifies the} eigenstrain inside any given inclusion $i$, \rev{so as}{} to account for the influence of the remaining inclusions $N\backslash i$. \rev{An iterative algorithm has been chosen because a closed form solution for multiple inclusions does not exist and a numerical solution would be prohibitively expensive~\cite{pichler2010estimation}}{}. 

\begin{figure}[!ht]
  \centering
  \includegraphics[width=70mm]{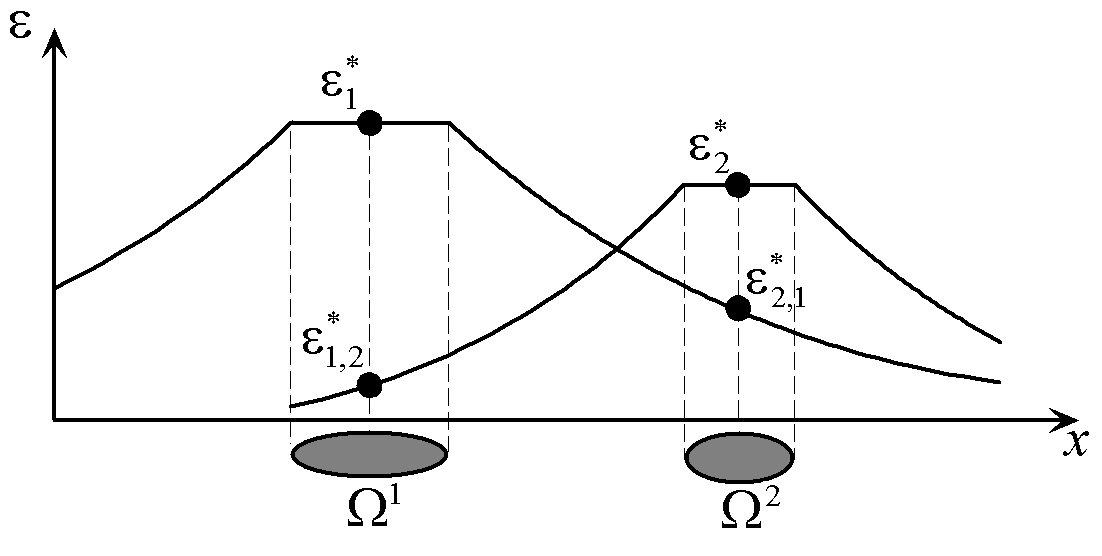}
  \caption{Principle of self-compatibility algorithm for double inclusion problem in 1D.}
  \label{fig:self-bal}
\end{figure}

First, the eigenstrain $\tenss{\strain}_i^\tau$ for each inclusion $i$ is calculated (Eq.~\ref{eqA:eq_incl_meth_step_5}) without reference to the other inclusions (\lline{2}). Next, the associated perturbation strain $\tenss{\strain}^*_i$ for each inclusion $i$ is evaluated (Eq.~\ref{eshelby}) at the centre of all other inclusions (\lline{3}).
The mutual interaction of inclusions is then taken into account via a correction
of the eigenstrain ($\Delta\tenss{\strain}^\tau_i$). For each inclusion $i$,
this correction is calculated (\lline{8}) from the inverse of the inclusion's
\aname{Eshelby} tensor $\tensf{S}_i^{-1}$ and the perturbation strains of all
other inclusions, evaluated at the centroid of inclusion $i$. The perturbation
strain resulting from inclusion $j$ at the centre of inclusion $i$ is denoted as
$\tenss{\strain}^*_{i,j}$. This is demonstrated in \Fref{fig:self-bal} for a two
inclusion problem in 1D. The eigenstrain correction \rev{$\Delta\tenss{\strain}^\tau_i$}{} is then used to calculate 
the correction to the perturbation strains (\lline{10}).\rev{}{ Finally, the
perturbation strain is updated (\lline{12}).} The algorithm continues until a
small \myemph{Euclidean} norm between the last two iterations of the total
eigenstrains is achieved. At convergence, the
corresponding stress and displacement perturbations are recalculated from
the corrected transformation eigenstrains.

\rev{
It is worthwhile noting that the algorithm does not depend on a particular sequence of inclusions, as follows from the elastic reciprocity theorem~\cite[and references therein]{pichler2010estimation}. Moreover, since the perturbation fields are calculated for the entire macroscopic domain (no RVE is considered), stress admissibility and strain compatibility, in the sense of macro-micro field relations, are fulfilled a priori.
}{}

The computational complexity of the \rev{Self-compatibility}{} algorithm is $O(N^2)$. However, this can be
improved by taking into account only those inclusions which have a
non-negligible influence to the inclusion of interest $i$. \rev{Preliminary studies have shown this to give a significant computational speed-up and will be reported in a future paper.}{}
\begin{table}[!ht]
  \begin{algorithm}{Self Compatibility Algorithm $(\tenss{\strain}^0_i, \tensf{B}_i, \tensf{S}_i, \tensf{S}^{-1}_i, N)$}
    \algline{\function{For} $(i \leq N)$}

    \algline{\algstep $\tenss{\strain}^\tau_i = \tensf{B}_i\dcontr\tenss{\strain}^0_i$ \quad(Eq.~\ref{eqA:eq_incl_meth_step_5})}

    \algline{\algstep $\tenss{\strain}^*_i = \tensf{S}_i\dcontr \tenss{\strain}^\tau_i$ \quad(Eq.~\ref{eshelby})}

    \algline{\algstep \function{Set}$\Delta\tenss{\strain}^*_i = \tenss{\strain}^*_i$}

    \algline{\function{EndFor}}

    \algline{\function{Do}}

    \algline{\algstep \function{For} $(i \leq N)$}

    \algline{\algstep \algstep $\Delta\tenss{\strain}^\tau_i = \sum_{j\backslash i}^N\tensf{S}^{-1}_i\dcontr \Delta\tenss{\strain}^*_{i,j}$ \quad(Eq.~\ref{eshelby})}


    \algline{\algstep \algstep $\tenss{\strain}^\tau_i = \tenss{\strain}^\tau_i + \Delta\tenss{\strain}^\tau_i$}

    \algline{\algstep \algstep $\Delta\tenss{\strain}^*_i = \tensf{S}_i\dcontr \Delta\tenss{\strain}^\tau_i$}



    \algline{\algstep \function{EndFor}}

    \algline{\function{While} $\big(\sum_{i}^N\norm{\Delta\tenss{\strain}^*_{i}} > \eta \big)$}
  \end{algorithm}
  \caption{\mname{Self-compatibility} algorithm. Note, that $\eta$ stands for an acceptable tolerance.}
  \label{alg:selfBalancing}
\end{table}

\section{Construction of perturbation approximation functions for FEA}
\label{FEA_perturbation}
%
The above methodology can be utilised to formulate an enhanced Finite Element
formulation. The primary task is to determine appropriate approximation
functions for the mechanical perturbation fields $\vek{u}^*$,
$\tenss{\strain}^*$ and $\tenss{\stress}^*$ based on the analytical
micromechanics developed above and which can then augment the standard macroscopic field
approximations. It should be noted that the \aname{Voigt-Mandel} notation is
exclusively used in the forthcoming text.

The perturbation field approximation functions are determined \myemph{a priori}
as a linear combination of the perturbation fields evaluated analytically for
six load cases, with self-equilibrium enforced by means of the
\mname{self-compatibility} algorithm outlined above (\Tref{alg:selfBalancing}). Each
load case corresponds to a unit component of the macroscopic strain vector
$\sevek{\vstrain}^0_i, \; i = 1,\dots,6$ and the resulting analytically
determined stress, strain and displacement perturbation fields are arranged,
column-by-column into $\semtrx{\mathfrak{s}}^*_{6\times 6}$,
$\semtrx{\mathfrak{e}}^*_{6\times 6}$ and $\semtrx{\mathfrak{u}}^*_{3\times 6}$
matrices, respectively:

%

%
\begin{equation}
  \semtrx{\mathfrak{s}}^* = \left[\begin{array}{ccc} ^1\sevek{\vstress}^*&\dots&^6\sevek{\vstress}^*\end{array}\right]
  ,
  \semtrx{\mathfrak{e}}^* = \left[
  \begin{array}{ccc}
    ^1\sevek{\vstrain}^* &\dots &^6\sevek{\vstrain}^*
  \end{array}
  \right]
  \label{eq:calculated_pert_s_e}
\end{equation}
\begin{equation}
  \semtrx{\mathfrak{u}}^* = \left[
  \begin{array}{ccc}
    ^1\sevek{u}^* &\dots &^6\sevek{u}^*
  \end{array}
  \right]
  \label{eq:calculated_pert_u}
\end{equation}
where the left superscript refers to a specific load case, $1$ to $6$.


\subsection{Partition of unity method}
\label{PUM}

Partition of Unity (PU) Methods (for example the eXtended Finite Element Method)
extend the underlying basis functions used for interpolating the displacement
field by adding an appropriate set of additional functions.
Following~\cite{Melenk:1996:PUM,Babuska:1997:PUM} it has been shown
that the displacement field $\sevek{u}(\x)$ within an element can be
interpolated by
\begin{equation}
\sevek{u}(\x)
=
\sum_{i=1}^n
\left(
  \semtrx{N}^i(\x)\sevek{a}^i
  +
  \semtrx{N}^i\semtrx{N}_\gamma(\x)\sevek{b}^i
\right)
\end{equation}
where $n$ is the number of nodes per element,
$\semtrx{N}^i(\x)=N^i(\x)\semtrx{I}$ is the standard matrix of element
shape functions for node $i$, $\semtrx{I}$ is the identity matrix and
$\sevek{a}^i$ the standard displacement degrees of freedom at node $i$.
$\semtrx{N}_\gamma$ is a matrix containing the additional basis terms
and $\sevek{b}^i$ are the associated additional degrees of freedom at node $i$.
It is important to recognise that the element shape functions form a partition
of unity, i.e.
\begin{equation}
\sum_{i=1}^n N^i(\x)=1
\end{equation}
The six analytically derived displacement
perturbation functions contained in $\semtrx{\mathfrak{u}}^*$ can be used as the additional functions
$\semtrx{N}_\gamma$ to augment the standard basis functions. Thus
\begin{equation}
\semtrx{N}_\gamma(\x)
=
\left[\begin{array}{ccccccccc}
^1 u_1^*(\x) & \dots & ^6 u_1^*(\x) &
           0 & \dots &             0&
           0 & \dots &             0\\
           0 & \dots &             0&
^1 u_2^*(\x) & \dots & ^6 u_2^*(\x)&
           0 & \dots &             0\\
           0 & \dots &             0&
           0 & \dots &             0&
^1 u_3^*(\x) & \dots & ^6 u_3^*(\x)
\end{array}\right]
\label{eq:N_gamma}
\end{equation}

With this at hand, the PU-based finite element formulation can be derived, see
for example~\cite{Wells2001}. PU methods are particularly useful in problems
where the extension of the basis functions is introduced on a node by node
basis, so that additional degrees of freedom are only introduced at nodes where
the basis is extended. One obvious example of such a local feature that can be
modelled in this way is discrete cracks~\cite{Wells2001}. However, this
favourable property is not exploited here because we wish to model a large
number of heterogeneities throughout the domain. In 3D problems, there are 3
standard displacement degrees of freedom per node; this would be extended by an
additional 18 degrees of freedom per node \rev{and per heterogeneity}{} with the proposed approach.

It is also worth noting that for standard finite elements, the volume
integration of the discrete system of equations is relatively straightforward.
However, extension of the basis functions to include the perturbation functions
in \Eref{eq:N_gamma} makes this process significantly more arduous.
For these reasons, an alternative Finite Element approach using Hybrid Trefftz Stress
elements~\cite{Freitas1998,Kaczmarczyk20091298} is considered where the standard
basis function is not extended, as with PU methods, but enhanced such that no
additional degrees of freedom are introduced. This is described in the next
section.

\section{Hybrid-Trefftz stress element formulation}\label{s:hts-formulation}
In this section a finite element formulation based on an enhancement of a
hybrid-Trefftz stress (HTS) element formulation~\cite{Kaczmarczyk20091298} is
presented.

%
\begin{figure}[!ht]
  \centering
  \includegraphics[width=70mm]{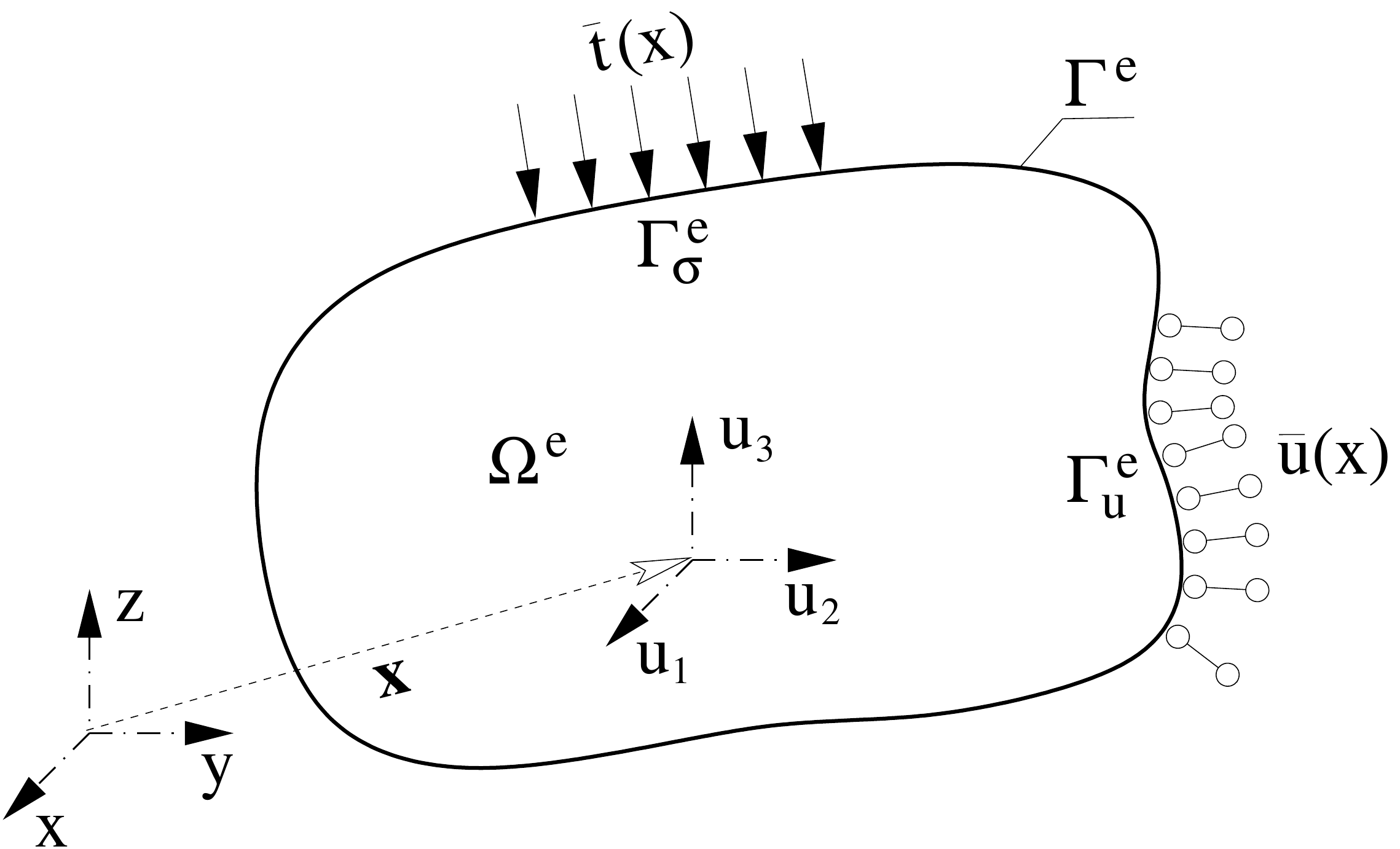}
  \caption{Elastic body representing HTS element}
  \label{fig:htse}
\end{figure}
The problem requires a solution to the displacement $\vek{u}$ and stress
$\tenss{\stress}$ fields as a result of given boundary displacements
$\overline{\vek{u}}$ and tractions $\overline{\vek{q}}$ on $\Gamma^e_u$ and
$\Gamma^e_\stress$, respectively. The displacement and stress fields must
fulfil the following governing equations:
\begin{equation}
  \begin{array}{llll}
    \semtrx{L}^\trn \sevek{\vstress} &= 0 &\textrm{in}~\Omega^e
    &\textrm{\dots Cauchy~equilibrium~equation}\\
    \semtrx{L}\sevek{u} &= \sevek{\vstrain} &\textrm{in}~\Omega^e
    &\textrm{\dots \rev{strain-displacement relationship}{}}\\
    \sevek{\vstress} &= \semtrx{C}\sevek{\vstrain}
    &\textrm{in}~\Omega^e &\textrm{\dots  constitutive~equation}\\
    \semtrx{N}\sevek{\vstress} &= \sevek{\overline{t}}
    &\textrm{on}~\Gamma^e_\stress
    &\textrm{\dots \rev{static~boundary~conditions}{}}\\
    \sevek{u} &= \sevek{\overline{u}} &\textrm{on}~\Gamma^e_u
    &\textrm{\dots \rev{kinematic~boundary~conditions}{}}
    \label{eq:governing_eqs}
  \end{array}
\end{equation}
where $\sevek{\vstress}$ and $\sevek{\vstrain}$ are the column matrix representation
of the second order stress and strain tensor, respectively, $\sevek{u}$ represents
the displacement vector, $\semtrx{C}$ is the matrix representation of
fourth order stiffness tensor and finally $\overline{\sevek{u}}$ and
$\overline{\sevek{t}}$ represent the applied displacements and tractions,
respectively. The gradient operator $\semtrx{L}$ and the matrix
of directional cosines $\semtrx{N}$ of the outward normal to element
boundary $\Gamma^e$ have the following forms~\cite{bittnar1996numerical}
\begin{equation}
  \begin{array}{cc}
    \semtrx{L} =
    \left[
      \begin{array}{ccc}
        \partial/\partial x & 0 & 0\\
        0 & \partial/\partial y & 0\\
        0 & 0 & \partial/\partial z\\
        \partial/\partial y & \partial/\partial x & 0 \\
        0 & \partial/\partial z & \partial/\partial y \\
        \partial/\partial z & 0 & \partial/\partial x
      \end{array}
      \right],
    &\semtrx{N} =
    \left[
      \begin{array}{cccccc}
        n_x &0 &0 &0 &n_z &n_y\\
        0 &n_y &0 &n_z &0 &n_x\\
        0 &0 &n_z &n_y &n_x &0
      \end{array}
      \right]
  \end{array}
\end{equation}
%

%
\subsection{Stress, strain and displacement approximations}
%
The macroscopic stress field within the HTS element is approximated as
%
%
%
\begin{equation}\label{eq:stress_approx}
\semtrx{\vstress}^0
=
\semtrx{S}_v^0
\sevek{v}
\mbox{ in }
\Omega^e
\end{equation}
where $\sevek{v}$ is the vector of \myemph{generalised} stress
degrees of freedom, $\semtrx{S}_v^0$ denotes the matrix of stress
approximation functions chosen so as to automatically satisfy the
equilibrium conditions \Eref{eq:governing_eqs}$^{1,4}$.
Thus,
\begin{equation}\label{eq:stress_approx_equilibrium}
\semtrx{L}^\trn\semtrx{S}_v^0 \sevek{v}
=
\sevek{0}
\mbox{ in }
\Omega^e
\end{equation}
and
\begin{equation}
\sevek{t}
=
\semtrx{N}
\semtrx{S}_v^0
\sevek{v}
\mbox{ on }
\Gamma^e
\end{equation}
where $\sevek{t}$ represents the traction vector induced by the
macroscopic stress approximation field.

The macroscopic strain and displacement fields are expressed analogously to Eq.~(\ref{eq:stress_approx}) as
\begin{equation}
  \sevek{\vstrain}^0 = \semtrx{E}_v^0
  \sevek{v}\quad\textrm{and}\quad\sevek{u}^0 = \semtrx{U}_v^0
  \sevek{v}
  \label{eq:strain_disp_approx}
\end{equation}
where $\semtrx{E}_v^{0}$ and $\semtrx{U}_v^{0}$ are directly associated with the stress approximation by
means of the compatibility equation \eqref{eq:governing_eqs}$^{2}$ and constitutive equation \eqref{eq:governing_eqs}$^{3}$ as
\begin{equation}
  \semtrx{S}_v^{0} = \semtrx{C}^0\semtrx{E}_v^{0} =
  \semtrx{C}^0\semtrx{L}\semtrx{U}_v^{0}
  \label{eq:const_eqs_approx}
\end{equation}
Since the stress approximation functions $\semtrx{S}_v^0$ are typically polynomial functions, the integration of $\semtrx{E}_v^0$ to get $\semtrx{U}_v^0$ is relatively straightforward.

Rather than extend the solution space to capture the influence of the
heterogeneities, as was briefly described in Section~\ref{PUM}, here we enhance
the macroscopic approximations to include the influence of the heterogeneities,
thereby not increasing the number of unknowns. The total stress (macroscopic
plus perturbation) field within the HTS element is approximated, following
Eq.~\eqref{eq:fields_decomposition}, as

\begin{equation}\label{eq:stress_approx2}
\semtrx{\vstress}
=
\semtrx{\vstress}^0
+
\semtrx{\vstress}^*
=
\left(
  \semtrx{S}_v^0 + \semtrx{S}_v^*
\right)
\sevek{v}
\mbox{ in }
\Omega^e
\end{equation}
where $\semtrx{S}_v^*$ is the perturbation counterpart to $\semtrx{S}_v^0$.
$\semtrx{S}_v^*$ can be constructed from \Eref{eq:calculated_pert_s_e}:
\begin{equation}
\semtrx{\vstress}^*=\semtrx{\mathfrak{s}}^*\semtrx{\vstrain}^0
\label{eq:stress_approx3}
\end{equation}
where $\semtrx{\mathfrak{s}}^*$ is the set of analytically defined stress perturbations for six load cases, each one representing a unit component of the macroscopic strain Eq.~\eqref{eq:calculated_pert_s_e}.
For the purposes of constructing $\semtrx{S}_v^*$, we approximate the macroscopic strain field as constant within each finite element and computed as the volume average of the actual macroscopic strain field. From Eq.~\eqref{eq:strain_disp_approx},
\begin{equation}
\semtrx{\vstrain}^{0,\textrm{ave}} = \frac{1}{\abs{\Omega^e}}\int_{\Omega^e}
  \semtrx{\vstrain}^0 \de{\Omega^e}= \frac{1}{\abs{\Omega^e}}\int_{\Omega^e}
  \semtrx{E}_v \de{\Omega^e}\sevek{v}= \semtrx{E}_v^{\textrm{ave}} \sevek{v}
\end{equation}
Substituting $\semtrx{\vstrain}^{0,\textrm{ave}}$ for $\semtrx{\vstrain}^0$ into Eq.~\eqref{eq:stress_approx3} leads to
\begin{equation}
\semtrx{\vstress}^*=\semtrx{\mathfrak{s}}^*\semtrx{E}_v^{\textrm{ave}} \sevek{v}
\end{equation}
Thus, the matrix of stress perturbation approximation functions is:
\begin{equation}
\semtrx{S}_v^*=\semtrx{\mathfrak{s}}^*\semtrx{E}_v^{\textrm{ave}}
\end{equation}
Analogously, the approximation of total strain and displacement
fields within the element domain is given by
\begin{equation}
  \sevek{\vstrain} = \left(\semtrx{E}_v^0 + \semtrx{E}_v^*\right)
  \sevek{v}\quad\textrm{and}\quad\sevek{u} = \left(\semtrx{U}_v^0 + \semtrx{U}_v^*\right)
  \sevek{v}
\end{equation}
where the perturbation approximation matrices $\semtrx{U}_v^*$ and
$\semtrx{E}_v^*$ are, as with their macroscopic counterparts, directly
associated with the stress approximation as
\begin{equation}
  \semtrx{S}_v^{*} = \semtrx{C}^0\semtrx{E}_v^{*} =
  \semtrx{C}^0\semtrx{L}\semtrx{U}_v^{*}
  \label{eq:const_eqs_approx2}
\end{equation}
It is worthwhile noting
that the stress perturbation fields and the corresponding traction perturbation fields,
approximated as $\semtrx{\vstress}^*=\semtrx{S}_v^* \sevek{v}$ and
$\sevek{t}^* = \semtrx{N}\semtrx{S}_v^*\sevek{v}$, remain in
self-equilibrium.

%
%
%
\subsection{Static boundary conditions}
%
Contrary to general condition in Eq.~\eqref{eq:governing_eqs}$^4$, the
equilibrium on the element traction boundary is imposed only in the
weighted residual sense as:
\begin{equation}
  \int_{\Gamma^e_\stress}
  \semtrx{W}_1^\trn\left[\semtrx{N}\left(\sevek{\vstress}^0 +
    \sevek{\vstress}^*\right) - \sevek{\overline{t}}\right] \de{\Gamma^e}
  = 0
  \label{eq:residual_cauchy_eq_on_boundary}
\end{equation}
along with $\semtrx{W_1}$ representing the matrix of weighting
functions. Replacing the total stress field in
Eq.~\eqref{eq:residual_cauchy_eq_on_boundary} by its approximation from
Eq.~\eqref{eq:stress_approx2}, the traction boundary condition becomes
\begin{equation}
  \int_{\Gamma^e_\stress}
  \semtrx{W}_1^\trn\semtrx{N}\left(\semtrx{S}_v^0 + \semtrx{S}_v^*\right)
  \sevek{v} \de{\Gamma^e} =
  \int_{\Gamma^e_\stress}
  \semtrx{W}_1^\trn\sevek{\overline{t}}\de{\Gamma^e}
  \label{eq:traction_BC}
\end{equation}
%
%
%
\subsection{Kinematic boundary conditions}
%
Compatibility
inside the element domain $\Omega^e$ is also enforced in a weighted residual
sense, such that:
\begin{equation}
  \int_{\Omega^e}
  \semtrx{W}_2^\trn\left(\sevek{\vstrain}^0 + \sevek{\vstrain}^* -
    \semtrx{L}\sevek{u}\right)
  \de{\Omega^e} = 0
  \label{eq:residual_kinematic_eq_on_boundary}
\end{equation}
Next, utilising integration by parts and applying Green's
theorem to $\semtrx{W}_2^\trn\semtrx{L}\sevek{u}$,
\Eref{eq:residual_kinematic_eq_on_boundary} results in
\begin{eqnarray}
  \int_{\Omega^e}
  \semtrx{W}_2^\trn\left(\semtrx{E}_v^0 + \semtrx{E}_v^*\right)\sevek{v}
  \de{\Omega^e} &+&
  \int_{\Omega^e} (\semtrx{L}^\trn\semtrx{W}_2)^\trn \sevek{u}
  \de{\Omega^e}\nonumber\\
  &-& \int_{\Gamma^e_\stress}
  \left(\semtrx{N} \semtrx{W}_2 \right)^\trn \sevek{u}\de{\Gamma^e} =
  \int_{\Gamma^e_u}
  \left(\semtrx{N} \semtrx{W}_2 \right)^\trn \sevek{\overline{u}}
  \de{\Gamma^e}
  \label{eq:Clapeyrons_theorem}
\end{eqnarray}
With the current formulation, it is not necessarily possible to find a solution
to both Eqs.~\eqref{eq:traction_BC} \& \eqref{eq:Clapeyrons_theorem} that
satisfies both traction and kinematic boundary conditions acting on the element
boundary. As a consequence, Eq.~\eqref{eq:Clapeyrons_theorem} is relaxed by
introducing an additional and independent approximation of displacements on the
element traction boundary:
%
%
%
\begin{equation}\label{eq:u_gamma_approx}
\sevek{u}_\Gamma
=
\semtrx{U}_\Gamma
\sevek{q}
\mbox{ in }
\Gamma^e_\stress
\end{equation}
\rev{Here,}{where} $\sevek{q}$ stands for the set of displacement unknowns
and
$\semtrx{U}_\Gamma$ is the matrix of boundary displacement approximation
functions. Such a formulation of the stress element leads to a hybrid approach
~\cite{Kaczmarczyk20091298,Freitas1998}.

Given the above consideration, introducing
\Eref{eq:u_gamma_approx} into \Eref{eq:Clapeyrons_theorem} results in
\begin{eqnarray}
  \int_{\Omega^e}
  \semtrx{W}_2^\trn\left(\semtrx{E}_v^0 + \semtrx{E}_v^*\right)\sevek{v}
  \de{\Omega^e} &+&
  \int_{\Omega^e} (\semtrx{L}^\trn\semtrx{W}_2 )^\trn\sevek{u}
  \de{\Omega^e}\nonumber\\
  &-& \int_{\Gamma^e_\stress}
  \left(\semtrx{N} \semtrx{W}_2 \right)^\trn \sevek{u}_\Gamma
  \de{\Gamma^e} = \int_{\Gamma^e_u}
  \left(\semtrx{N} \semtrx{W}_2 \right)^\trn \sevek{\overline{u}}
  \de{\Gamma^e}
  \label{eq:Clapeyrons_theorem_u_gamma}
\end{eqnarray}
%
%
\subsection{Weighting functions}
%
In order to achieve an energy-consistent formulation, it is required
that all weighted terms within the integrals defined above have the dimension of work. The \myemph{weighting functions}
then directly follow from the integrands in
\Eref{eq:residual_cauchy_eq_on_boundary} and
\Eref{eq:residual_kinematic_eq_on_boundary} representing the increment
of internal work of strains within $\Omega^e$ and external work of
tractions on $\Gamma^e_\stress$, respectively. These functions thus
admit the following forms:
\begin{eqnarray}\label{eq:weight_fce_w}
\semtrx{W}_1
=
\semtrx{U}_\Gamma,
&&
\semtrx{W}_2
=
\semtrx{S}^0_v
\end{eqnarray}

First, introducing \Eref{eq:weight_fce_w}$^2$ into
\Eref{eq:Clapeyrons_theorem_u_gamma} and taking into account
condition~\eqref{eq:stress_approx_equilibrium} yields
\begin{equation}
  \int_{\Omega^e}\left(\semtrx{S}_v^0\right)^\trn \left(\semtrx{E}_v^0
  + \semtrx{E}_v^*\right)\sevek{v}\de{\Omega^e} -
  \int_{\Gamma^e_\stress}\left(\semtrx{N} \semtrx{S}_v^0\right)^\trn
  \semtrx{U}_\Gamma\sevek{q}\de{\Gamma^e} =
  \int_{\Gamma^e_u}\left(\semtrx{N} \semtrx{S}_v^0\right)^\trn
  \sevek{\overline{u}}\de{\Gamma^e}
  \label{eq:fist_eq_of_discrete_system}
\end{equation}
Second, introducing Eq.~\eqref{eq:weight_fce_w}$^1$
into the traction boundary condition~\eqref{eq:traction_BC} yields
\begin{equation}
  \int_{\Gamma^e_\stress}\semtrx{U}_\Gamma^\trn\semtrx{N}(\semtrx{S}_v^0+\semtrx{S}_v^*)
  \sevek{v}\de{\Gamma^e} =
  \int_{\Gamma^e_\stress}\semtrx{U}_\Gamma^\trn \sevek{\overline{t}}
  \de{\Gamma^e}
  \label{eq:second_eq_of_discrete_system}
\end{equation}
Combining Eqs.~\eqref{eq:fist_eq_of_discrete_system} \& \eqref{eq:second_eq_of_discrete_system} results in a coupled system of linear equations that can be expressed in compact form as
\begin{equation}
  \left[
    \begin{array}{cc}
      \semtrx{F} &-\semtrx{A}^\trn\\
      -(\semtrx{A}+\semtrx{A}^*) &\semtrx{0}
    \end{array}
    \right]
  \left[
    \begin{array}{c}
      \sevek{v}\\
      \sevek{q}
    \end{array}
    \right]
  =
  \left[
    \begin{array}{c}
      \sevek{p_u}\\
      \sevek{-p}_\stress
    \end{array}
    \right]
  \label{eq:HTS_elem_compact_SLAE}
\end{equation}
where the submatrices on the left-hand side follow from
Eqs (\ref{eq:fist_eq_of_discrete_system},~\ref{eq:second_eq_of_discrete_system})
and are given by the following integrals
\begin{eqnarray}
 \label{eq:Fmatrix}
  \semtrx{F}
  &=& \int_{\Omega^e}\left(\semtrx{S}_v^0\right)^\trn \left(\semtrx{E}_v^0
  + \semtrx{E}_v^*\right)\de{\Omega^e} =
  \int_{\Gamma^e} \semtrx{N}\left(\semtrx{S}_v^0\right)^\trn
  \left(\semtrx{U}_v^0 + \semtrx{U}_v^*\right) \de{\Gamma^e}\\
  \semtrx{A} &=&
  \int_{\Gamma^e_\stress}\semtrx{U}_\Gamma\semtrx{N}\semtrx{S}_v^0\de{\Gamma^e}\quad\textrm{and}\quad\semtrx{A}^* =
  \int_{\Gamma^e_\stress}\semtrx{U}_\Gamma\semtrx{N}\semtrx{S}_v^*\de{\Gamma^e}
  \label{eq:HTS_SLAE_LS}
\end{eqnarray}
and for the terms on the right-hand side it holds
\begin{equation}
  \sevek{p_u} = \int_{\Gamma^e_u}
  \left(\semtrx{N}\semtrx{S}_v^0\right)^\trn \sevek{\overline{u}}
  \de{\Gamma^e},
  \textrm{~and}\quad
  \sevek{p}_\stress = \int_{\Gamma^e_\stress}\semtrx{U}_\Gamma^\trn
  \sevek{\overline{t}} \de{\Gamma^e}
  \label{eq:HTS_SLAE_RS}
\end{equation}
Note that Eq.~\eqref{eq:Fmatrix} illustrates that the $\semtrx{F}$~matrix can be
evaluated via a boundary rather than volume integral. Thus all terms in
Eq.~\eqref{eq:HTS_elem_compact_SLAE} can be evaluated using boundary integrals
only.

The size of the system of equations to be solved simultaneously can be reduced
via static condensation, representing a significant reduction in computational
effort. First, from the first equation in Eq.~\eqref{eq:HTS_elem_compact_SLAE},
the generalised stress degrees of freedom $\sevek{v}$ are expressed in terms of
the displacement degrees of freedom $\sevek{q}$ as

\begin{equation}
       \sevek{v}=\semtrx{F}^{-1}(\sevek{p_u}+\semtrx{A}^\trn\sevek{q})
  \label{eq:stat_cond1}
\end{equation}
This is then substituted into the second equation of Eq.~\eqref{eq:HTS_elem_compact_SLAE} to yield:
\begin{equation}
       \left[(\semtrx{A}+\semtrx{A}^*)\semtrx{F}^{-1}\semtrx{A}^\trn\right]\sevek{q}= \sevek{p}_\stress-(\semtrx{A}+\semtrx{A}^*)\semtrx{F}^{-1} \sevek{p_u}
  \label{eq:sparse}
\end{equation}
This sparse system of equations is then solved for the displacement degrees of freedom $\sevek{q}$. Subsequently, the stress degrees of freedom $\sevek{v}$ can then be calculated on an element-by-element basis.

Our implementation of these HTS elements for composite materials (C-HTS elements) utilises displacement degrees of freedom that are associated with element faces rather than vertices. This has the advantage that the bandwidth of the stiffness matrix is minimised, as is interprocessor communication.

\section{Numerical Examples}\label{s:results}
%
The performance of the key components of the proposed
strategy (micromechanical solution, self-compatibility algorithm, finite element analysis convergence, etc.), in terms of efficiency and accuracy, have been explored through two numerical
examples.

\subsection{Three ellipsoidal inclusions in matrix}
\begin{figure}[!ht]
  \centering
  \includegraphics[height=4.2cm]{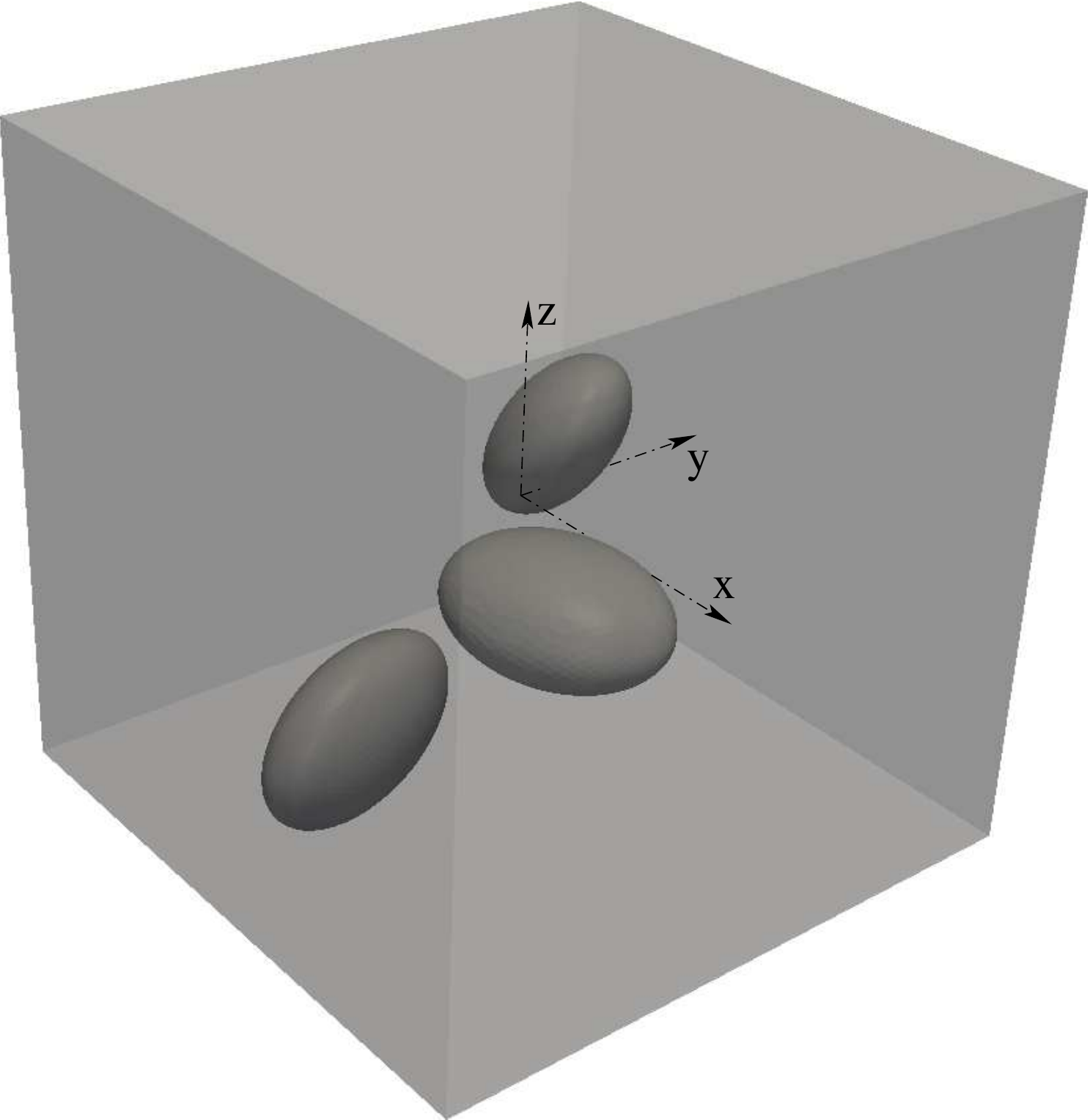}
  \caption{Geometry and topology of triple inclusion problem}
  \label{fig:3_incl_geometry}
\end{figure}

This example comprises three ellipsoidal inclusions embedded in a cube of matrix. The geometry of the problem is illustrated in~\Fref{fig:3_incl_geometry} and details of the ellipsoids are given in Table~\ref{tab:inclusionGeomParam}, including the semi-axes' dimensions, centroid coordinates and \aname{Euler} angles
$\phi$, $\nu$ and $\psi$, which are successive rotations of the semi-axes $a_1$, $a_2$ and $a_3$ about global coordinate axes $z$, $x$ and $z$, respectively. The cube has side lengths of $600$. The displacement boundary conditions were prescribed on faces $x = 300$, $y = 300$ and $z = 300$ as $\overline{u}_x = 0$, $\overline{u}_y = 0$ and $\overline{u}_z = 0$,
respectively. The remaining faces at $x = -300$, $y = -300$ and $z = -300$ were subject to uniform normal unit tractions. The Young's modulus for the homogeneous matrix was chosen as $E = 1$ and for the heterogeneities as $E = 2$. Poisson's ratio was chosen as $\nu = 0.1$ for both materials. All units are consistent.

It is worthwhile noting that the small contrast
in stiffness between the two materials was chosen deliberately to maximise the extent of the perturbation fields emanating from the heterogeneities. Large contrasts in stiffness between the matrix and heterogeneities lead to perturbation fields that decay rapidly with distance from the heterogeneities.
The close proximity of the three ellipsoidal heterogeneities to each other was also chosen deliberately in order to demonstrate the ability of the formulation to capture the interaction of multiple heterogeneities. Furthermore, the close proximity of one of the ellipsoids to a traction boundary was chosen to demonstrate the ability of the formulation to capture boundary effects.
%
%
%

%
\begin{table}[!ht]
  \centering
    \begin{tabular}{|c|c|c|c|c|c|c|c|c|c|}
      \hline
      Incl. & \multicolumn{3}{c|}{Centroid coordinates}
      &\multicolumn{3}{c|}{Semiaxes dim.}
      &\multicolumn{3}{c|}{Euler angles of $\max{a_i}$}\\
      \cline{2-10}
      &$x$ &$y$ &$z$ &$a_1$ &$a_2$ &$a_3$ &$\phi$ &$\nu$ &$\psi$\\
      \hline\hline
      1 &-48.07 &78.27 &14.81 &50 &75 &100 &74.21 &48.44 &-48.07\\
      \hline
      2 &16.45 &178.64 &-154.51 &50 &100 &75 &37.27 &22.27 &-25.51\\
      \hline
      3 &127.93 &-65.94 &-27.32 &100 &75 &50 &46.74 &11.17 &-26.30\\
      \hline
    \end{tabular}
    \caption{Topology and geometry of ellipsoidal inclusions of triple
      inclusion problem}
    \label{tab:inclusionGeomParam}
\end{table}

\begin{figure}[!ht]
  \centering
    \begin{tabular}{cc}
      \includegraphics[height=4.5cm]{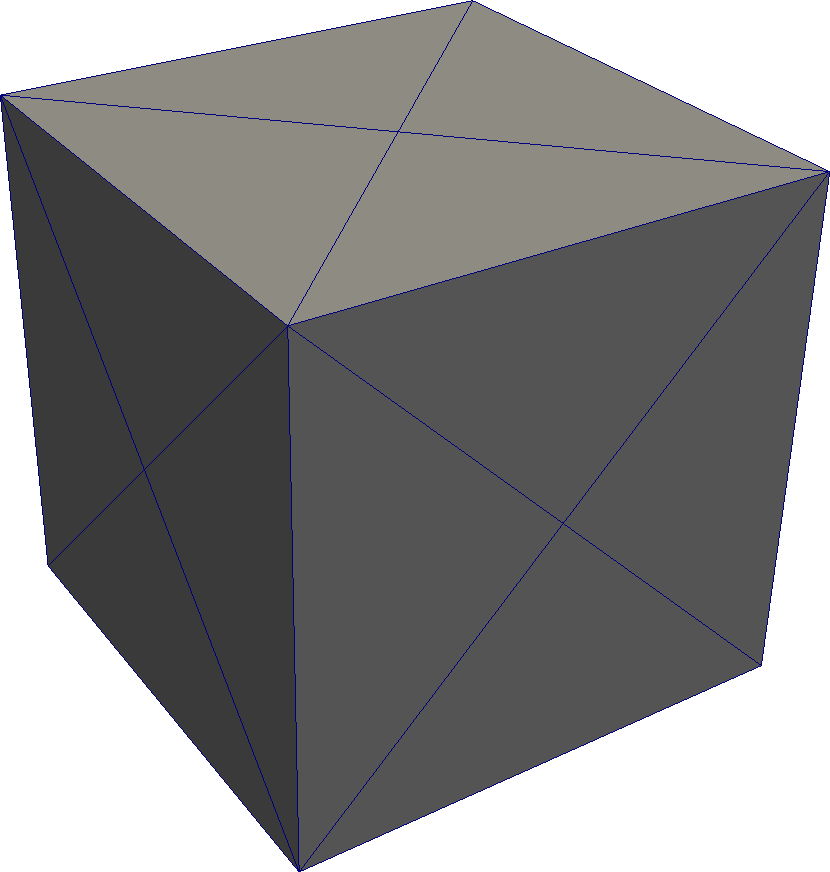}
    &\includegraphics[height=4.5cm]{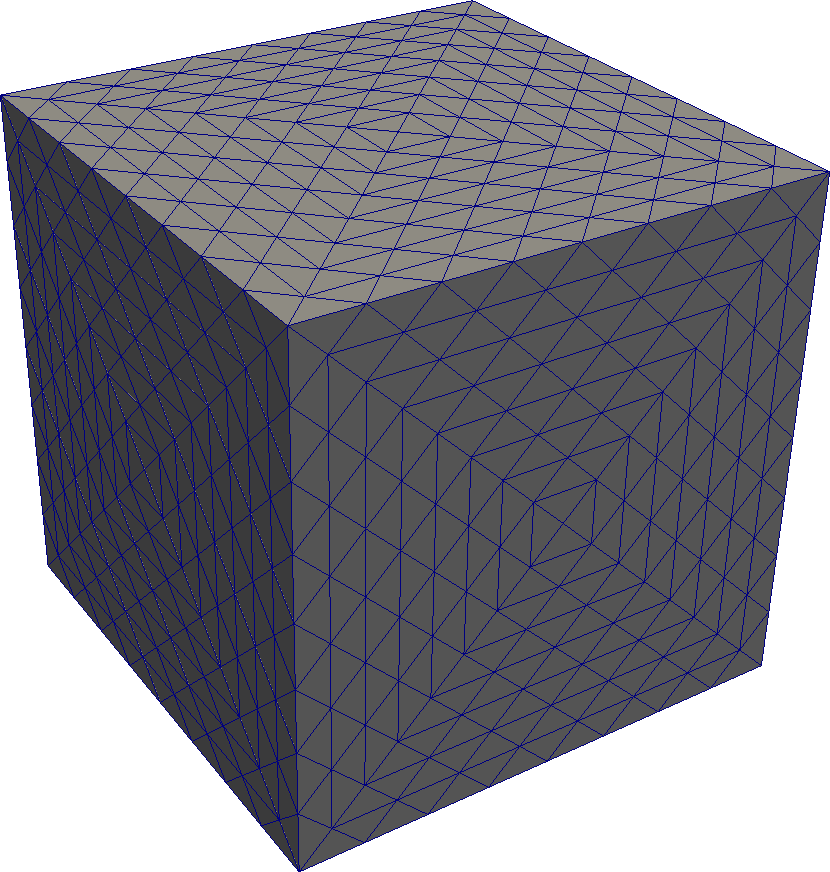}\\
    (a) &(b)
    \end{tabular}
    \caption{Triple inclusion task discretization by C-HTS elements:
      a) Coarse mesh with 24 enhanced elements (540 DOFs) b) Finer mesh with 1,536 enhanced elements
      (29,376 DOFs)}
    \label{fig:3d_CHTSE_meshes}
\end{figure}

\begin{figure}[!ht]
  \centering
    \begin{tabular}{ccc}
      \includegraphics[height=3.95cm]{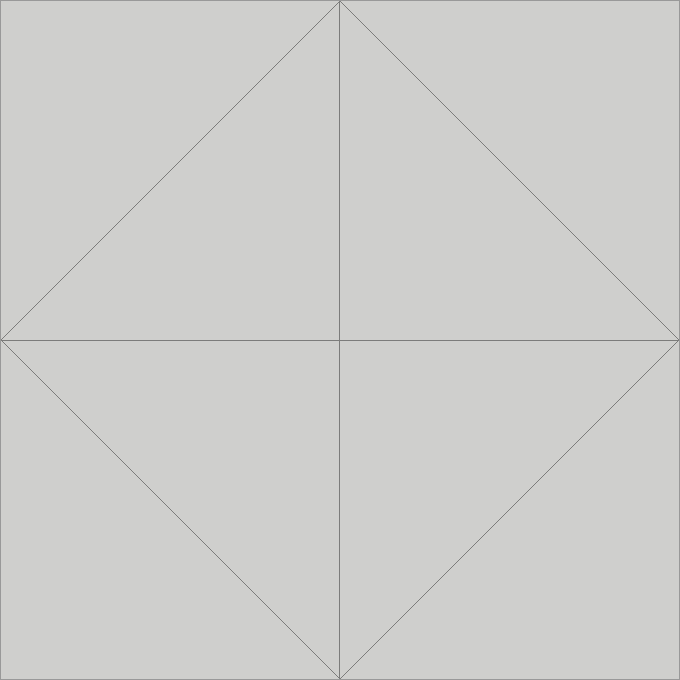}
    &\includegraphics[height=3.95cm]{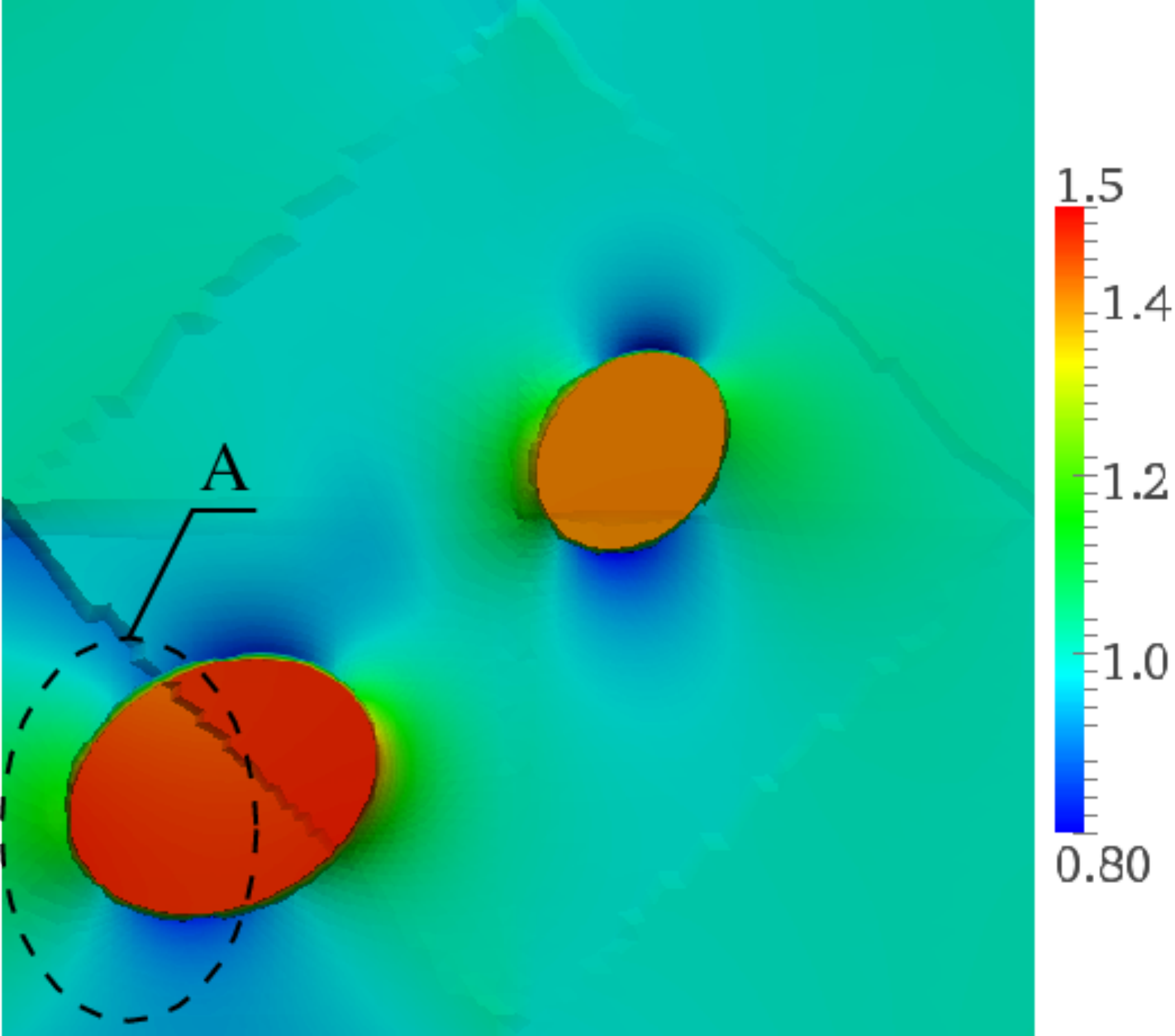}
    &\includegraphics[height=3.95cm]{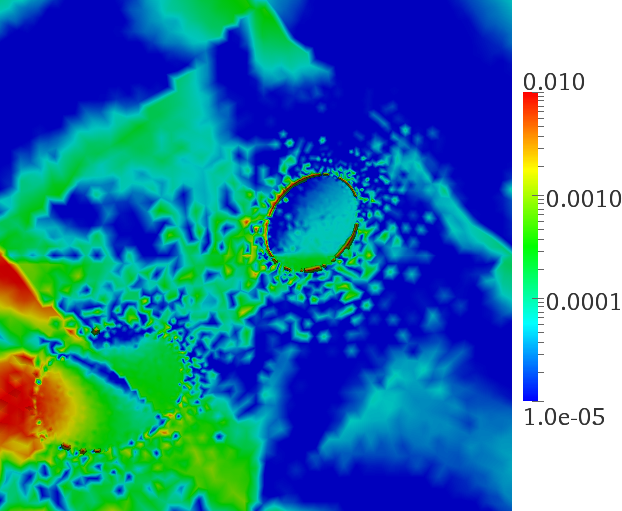}\\
    (a) &(b) &(c)
    \end{tabular}
    \caption{Coarse mesh solution: a) Enhanced finite element mesh
      in $yz$ plane at $x=0$, b) $\stress_{yy}$ in $yz$ plane,
      c) error calculated as $(\stress^{\textrm{ref}}_{yy} -
      \stress_{yy})^2/\stress^2_{yy}$}
    \label{fig:results_1}
\end{figure}

\begin{figure}[!ht]
  \centering
    \begin{tabular}{ccc}
      \includegraphics[height=3.95cm]{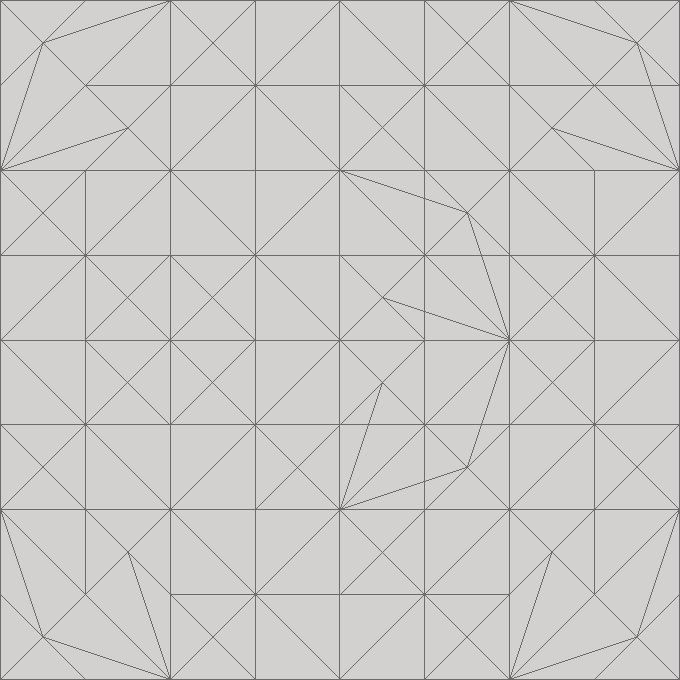}
    &\includegraphics[height=3.95cm]{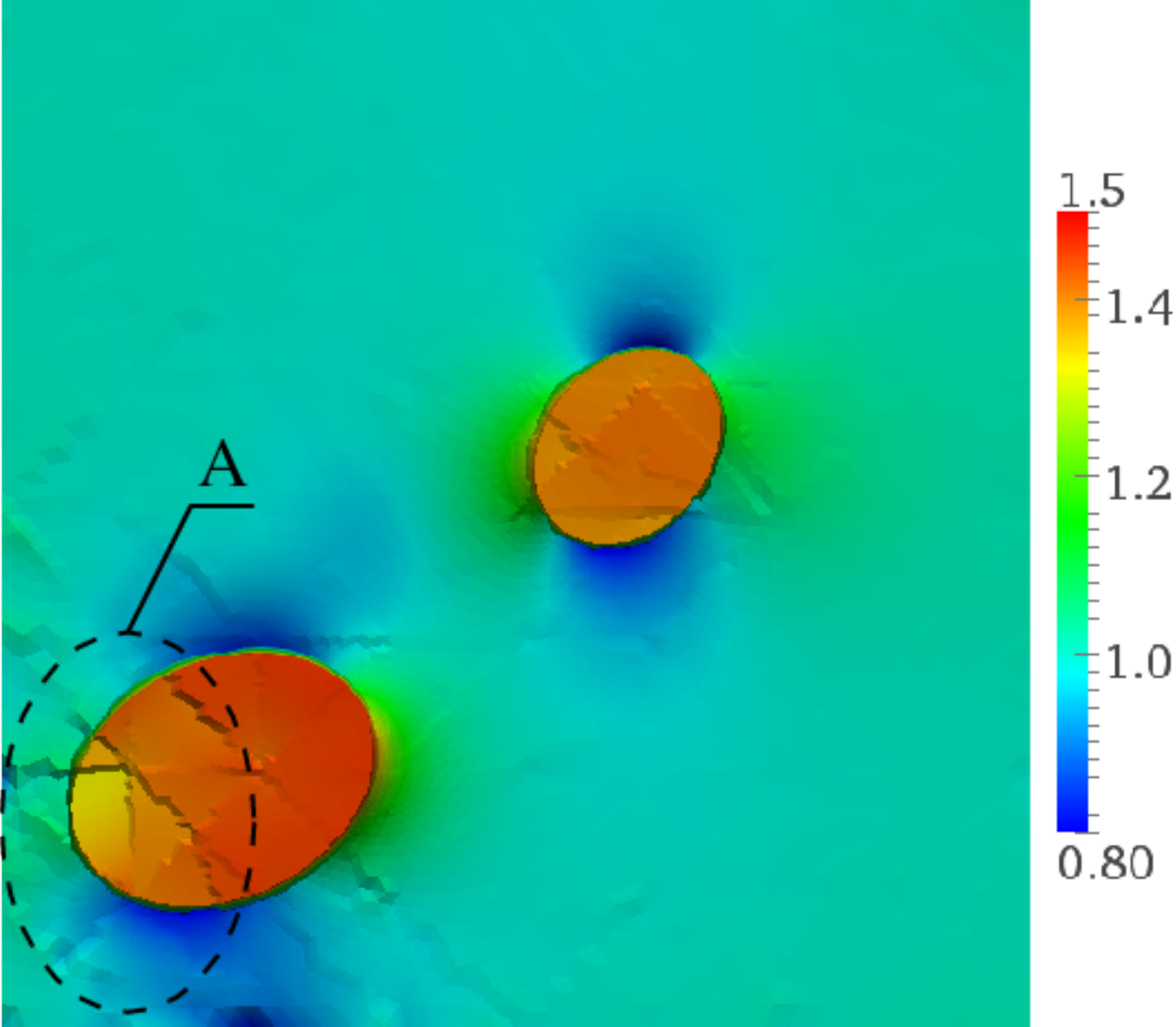}
    &\includegraphics[height=3.95cm]{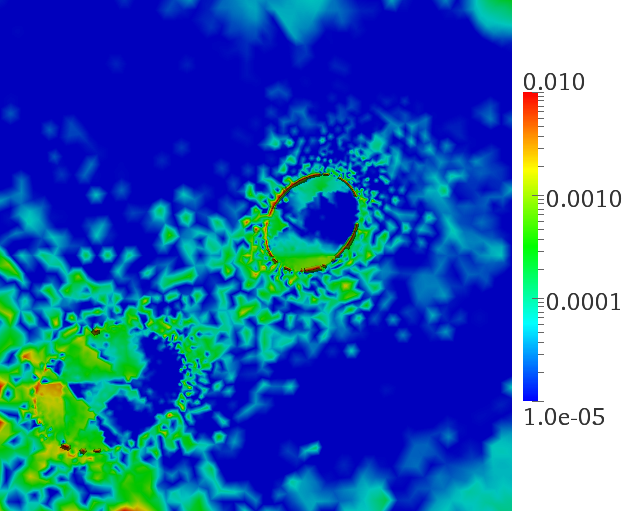}\\
    (a) &(b) &(c)
    \end{tabular}
    \caption{Finer mesh solution: a) Enhanced finite element mesh
      in $yz$ plane at $x=0$, b) $\stress_{yy}$ in $yz$ plane,
      c) error calculated as $(\stress^{\textrm{ref}}_{yy} -
      \stress_{yy})^2/\stress^2_{yy}$}
    \label{fig:results_2}
\end{figure}

The problem was analysed using two three-dimensional finite element meshes with different densities, comprising C-HTS elements. The coarse mesh comprised
$24$ elements and $540$~DOFs (\Fref{fig:3d_CHTSE_meshes}a) and the second,
refined, mesh comprised $192$ elements and $3888$~DOFs
(\Fref{fig:3d_CHTSE_meshes}b).
Results from the two enhanced finite element analyses are plotted in the $yz$-plane (at $x=0$). \Fref{fig:results_1}a and \Fref{fig:results_2}a show the two meshes in this plane. The $\stress_{yy}$ stress component for both analyses are shown in \Fref{fig:results_1}b and \Fref{fig:results_2}b.

\begin{figure}[!ht]
  \centering
    \begin{tabular}{cc}
    \includegraphics[height=4.2cm]{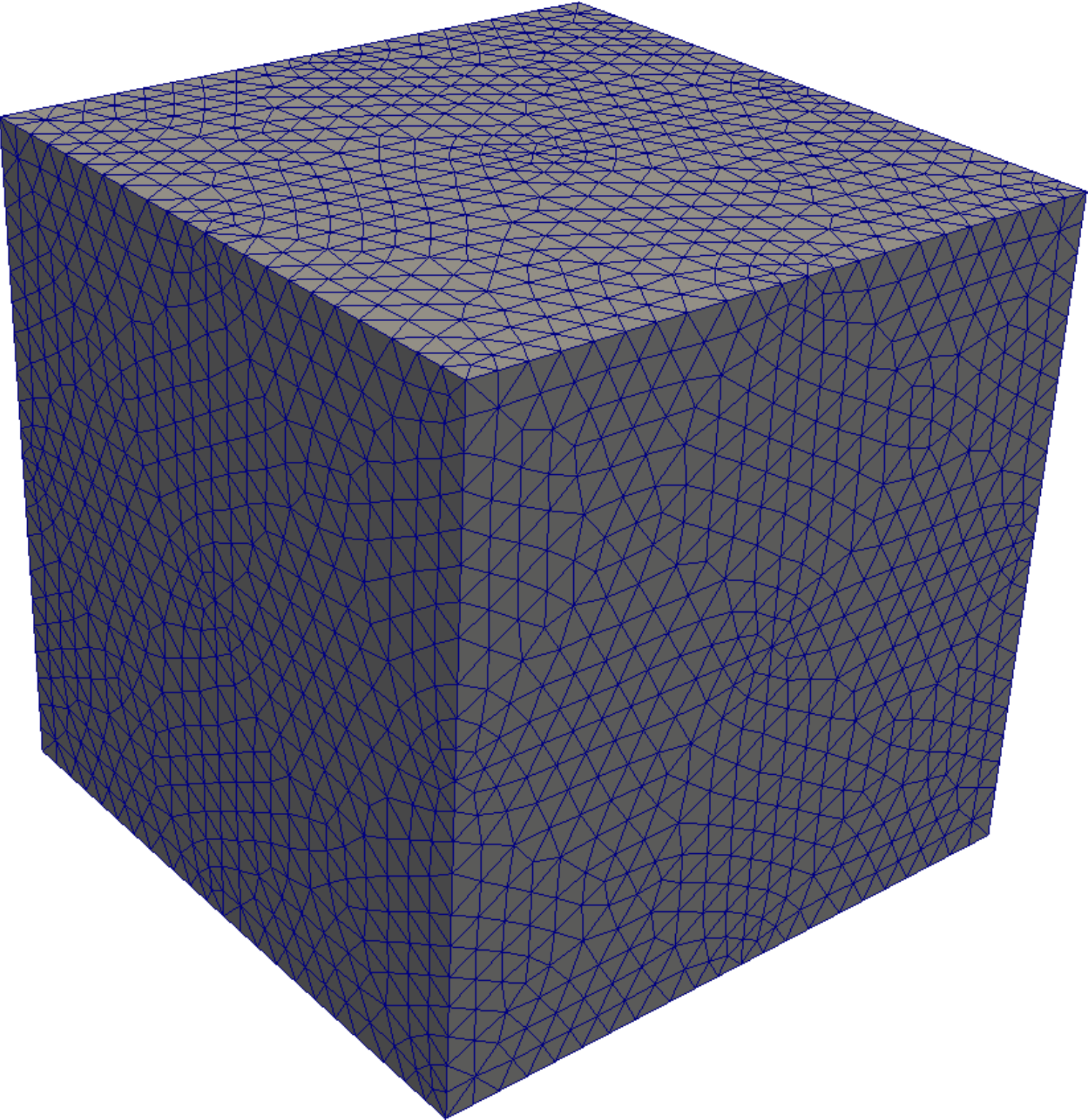}
    &\includegraphics[height=4.2cm]{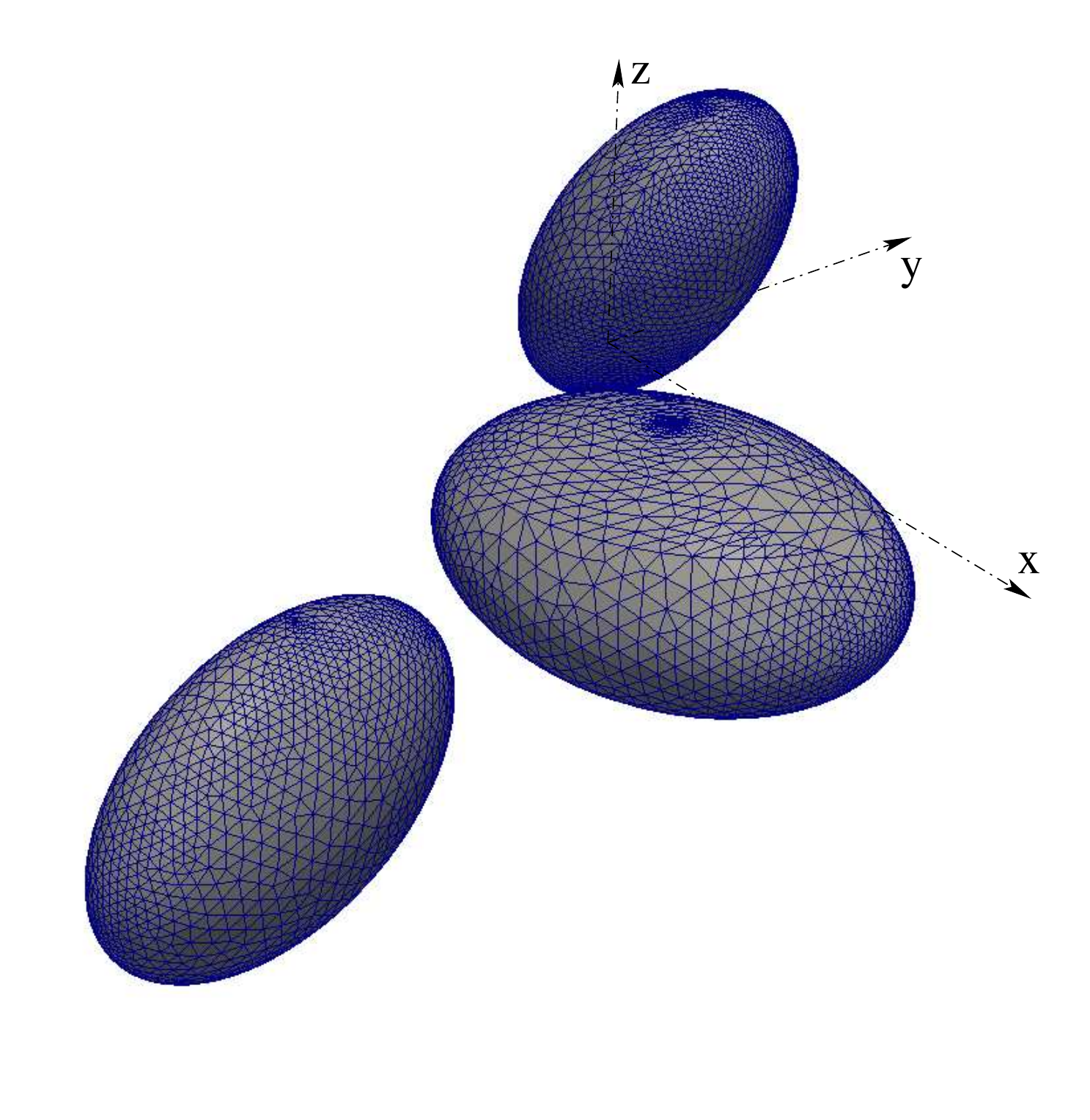}\\
    (a) &(b)\\*[2em]
      \includegraphics[height=4cm]{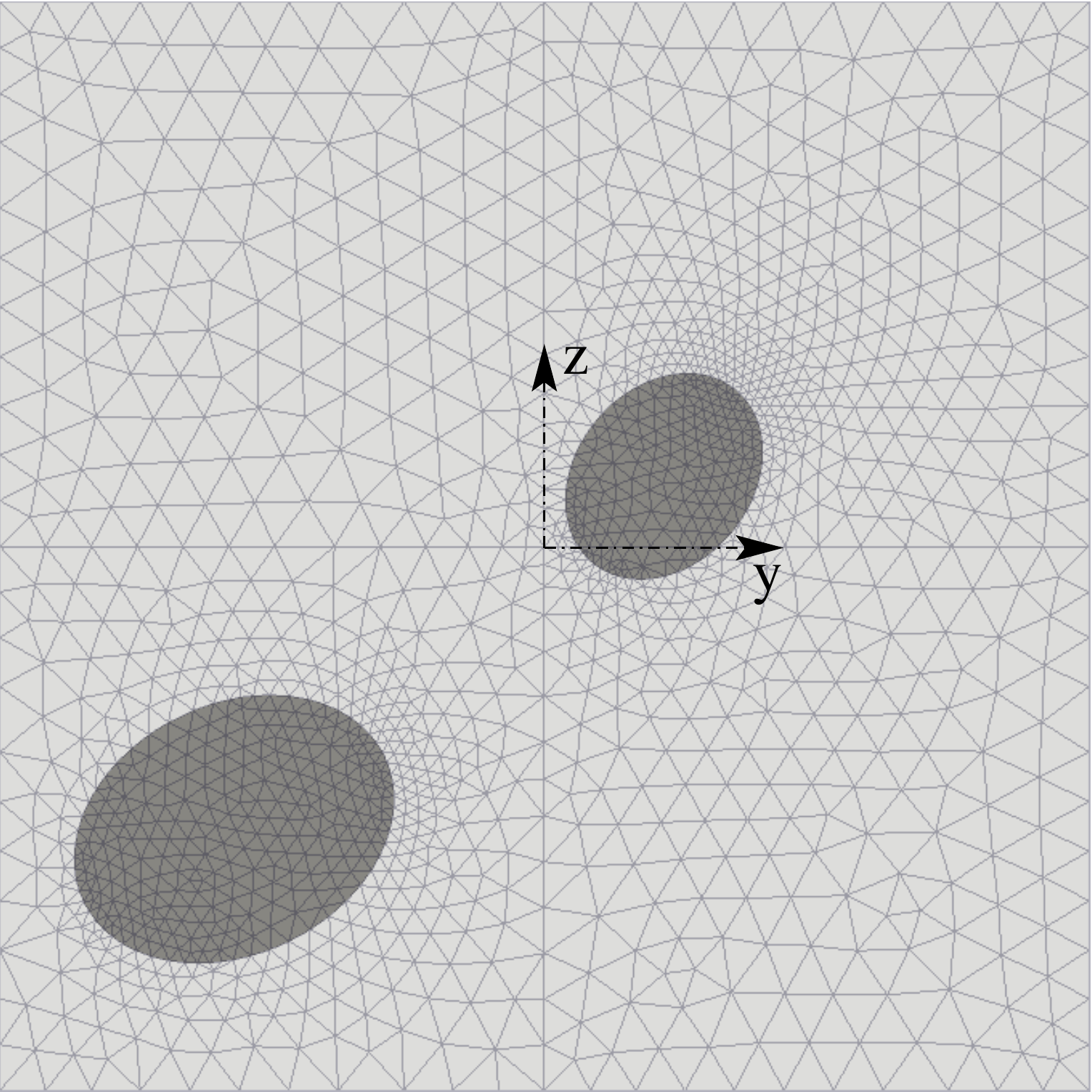}
    &\includegraphics[height=4cm]{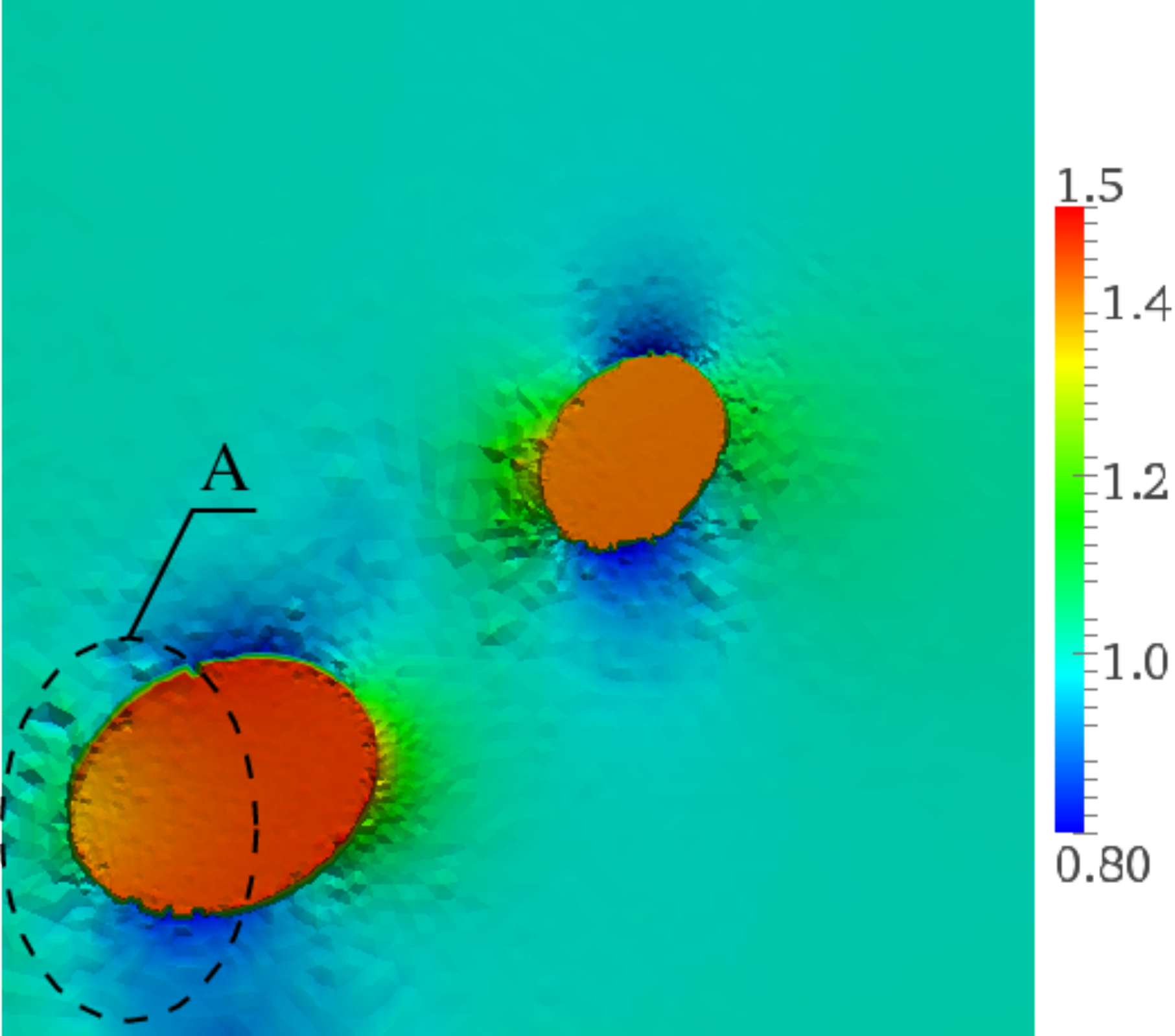}\\
    (c) &(d)
    \end{tabular}
    \caption{Reference analysis: a) discretization of entire body containing 309,406
      tetrahedra, b) mesh refinement on surface of heterogeneities, c) Reference finite
      element mesh in $yz$ plane at $x=0$, d) $\stress^{\textrm{ref}}_{yy}$ in $yz$ plane.}
    \label{fig:3_incl_reference_discretization}
\end{figure}

In addition, a reference finite element analysis of the same problem was undertaken for comparison sake. The reference analysis utilised HTS elements but without the proposed enhancement. Unlike
the other two analyses, the reference analysis utilised a mesh that explicitly resolved the three ellipsoidal heterogeneities and comprised $309,406$ tetrahedrons with $5,596,776$~DOFs
(\Fref{fig:3_incl_reference_discretization}a). The corresponding mesh and stress results of the reference analysis are shown in \Fref{fig:3_incl_reference_discretization}. Comparison of the stress results from the enhanced formulation and the reference analysis leads to the relative error plots shown in
\Fref{fig:results_1}c and \Fref{fig:results_2}c. It can be seen that even the very coarse mesh with the enhanced formulation results in good agreement.

\begin{figure}[!ht]
  \centering
    \begin{tabular}{cc}
      \includegraphics[height=3.5cm]{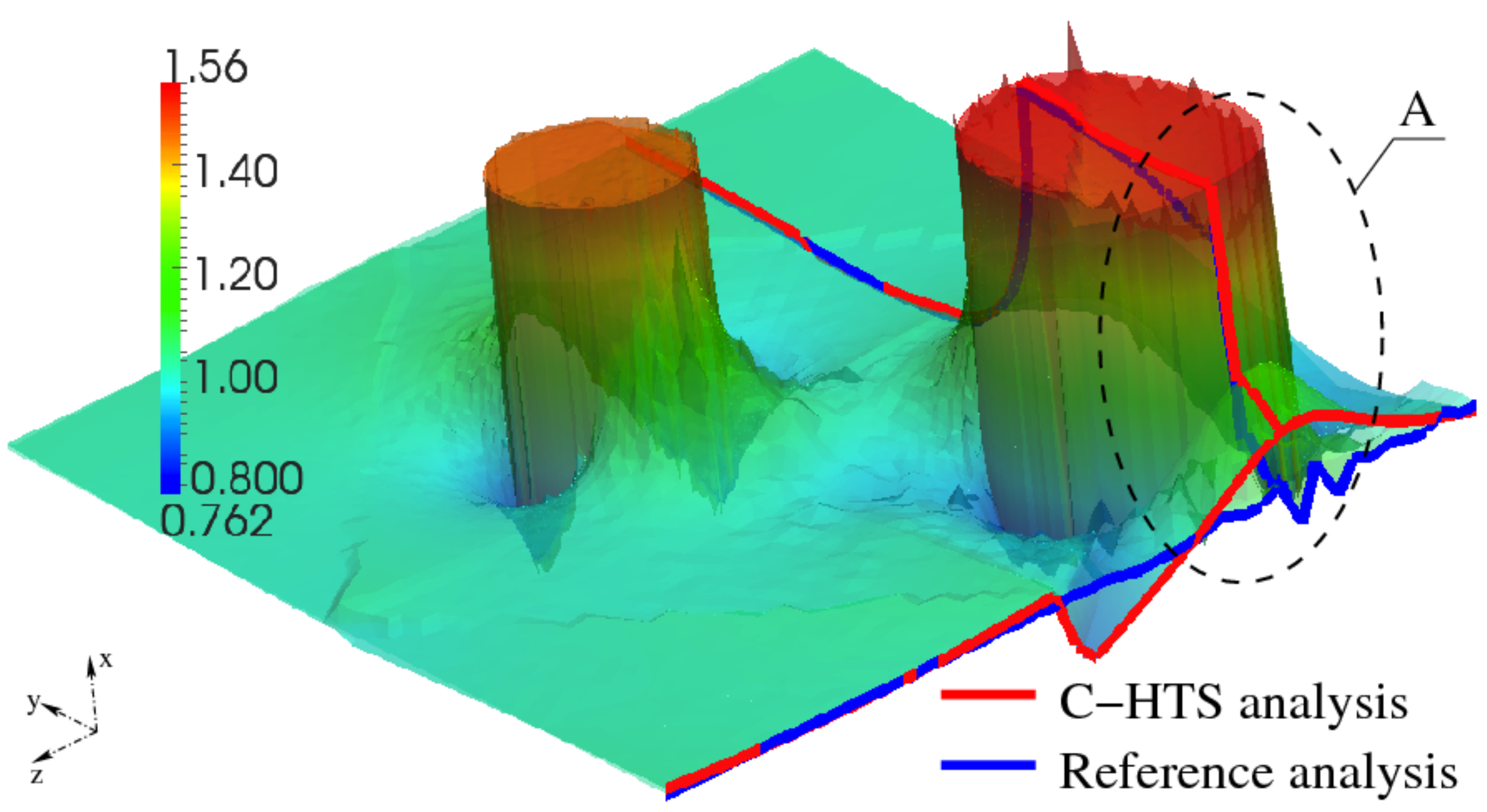}
    &\includegraphics[height=3.5cm]{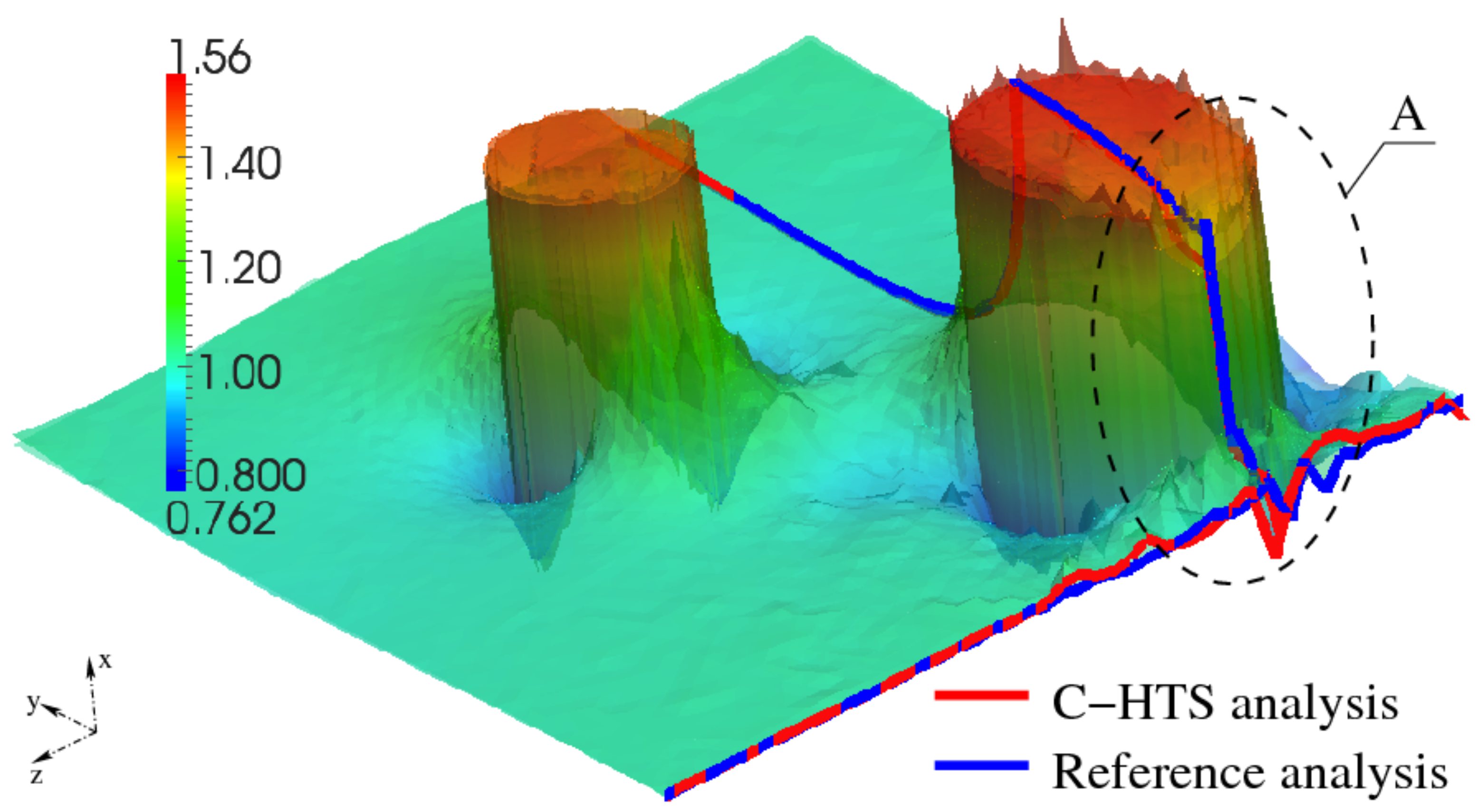}\\
    (a) &(b)
    \end{tabular}
    \caption{Detailed $3D$-plots of $\sigma_{yy}$ stress concentrations due to the
      boundary effects in
      $yz$ plane at $x=0$: Comparison of solution from reference analysis with a) coarse mesh C-HTS solution with 24 enhanced elements b)
      refined mesh C-HTS solution with 1,536 enhanced elements}
    \label{fig:3d_plots}
\end{figure}

Further comparison of the stress results is shown in \Fref{fig:3d_plots}, where it can been seen that along the traction boundary, the finer mesh of enhanced elements is able to capture the imposed constant stress field more accurately than the very much finer reference mesh.
The perturbation fields are based on the assumption of a heterogeneity in an infinite medium but the enhanced formulation still exhibits convergence in the regions strongly influenced by the traction boundary.
\subsection{L-shaped specimen}

\begin{figure}[!ht]
  \centering
    \begin{tabular}{cc}
    \includegraphics[height=4.5cm]{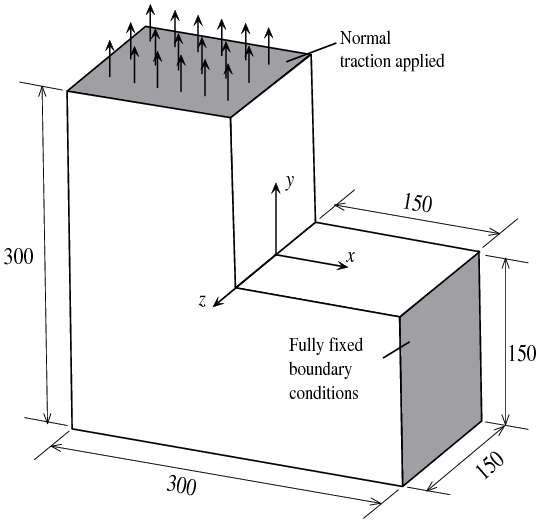}
    &\includegraphics[height=4.5cm]{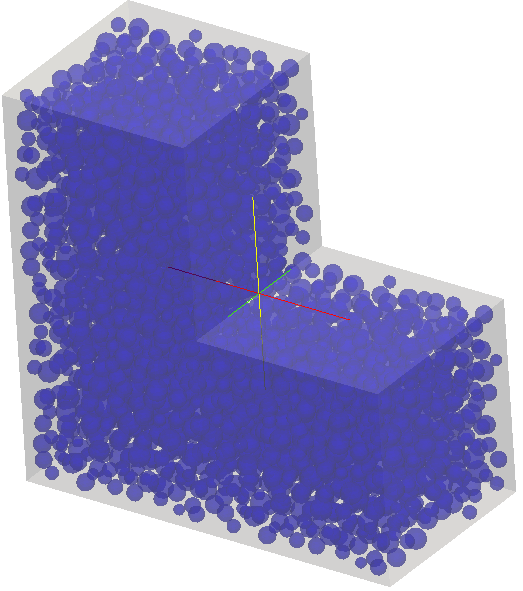}\\
    (a) &(b)
    \end{tabular}
    \begin{tabular}{ccc}
    \includegraphics[height=4.5cm]{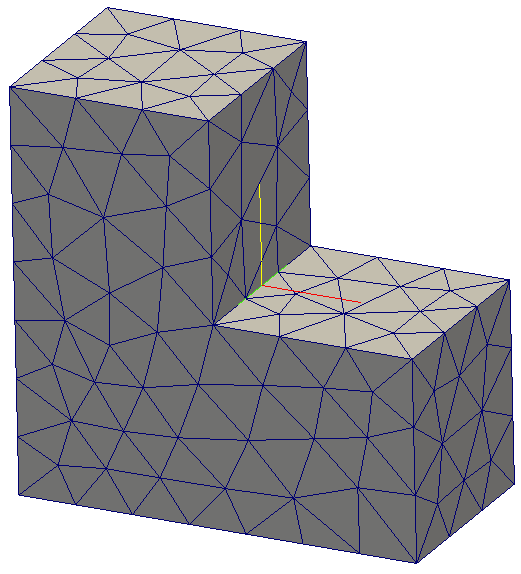}
    &\includegraphics[height=4.5cm]{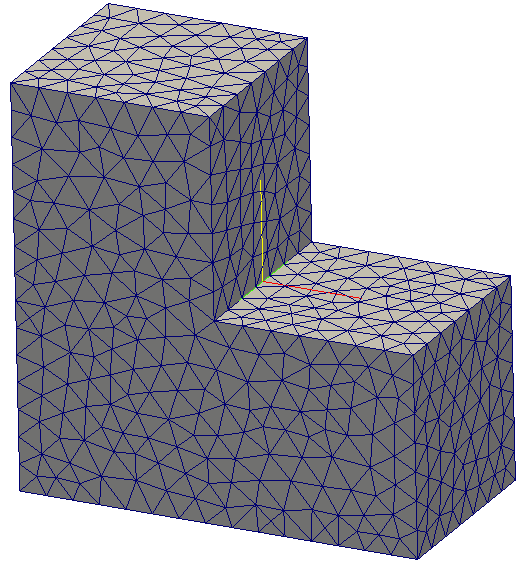}
    &\includegraphics[height=4.5cm]{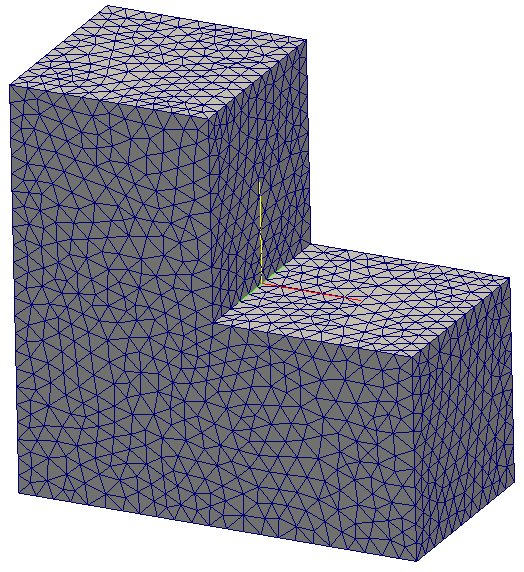}\\
    (c) &(d) &(e)
    \end{tabular}
    \caption{L-shaped specimen a) geometry, b) microstructure comprising 2,523 spherical inclusions,
  c) coarse mesh comprising 1074 tetrahedral elements,
  d) medium mesh comprising 8772 elements and
  e) fine mesh comprising 30007 elements.
\label{fig:l_shape1}}
\end{figure}

The proposed modelling strategy is also demonstrated on an example with a large number of
inclusion. A 3D L-shaped specimen with fully fixed boundary conditions on the right surface of the right-hand arm
and normal traction applied on the top surface of the left-hand arm is analysed, see Fig.~\ref{fig:l_shape1}. The length of the
plate is $300$ in both $x$ and $y$ direction, the depth is $150$ in $z$
direction. The Young moduli were chosen as $E=1$
and $E=2$ for matrix and inclusion respectively. Poisson's ratio was $\nu = 0.1$ for both
material phases. The microstructure comprised 2,523 spherical inclusions varying in size between $4$ and $8$ with a uniform spatial distribution (\Fref{fig:l_shape1}b). All units are consistent.

\begin{figure}[!ht]
  \centering
    \begin{tabular}{cc}
    \includegraphics[height=7cm]{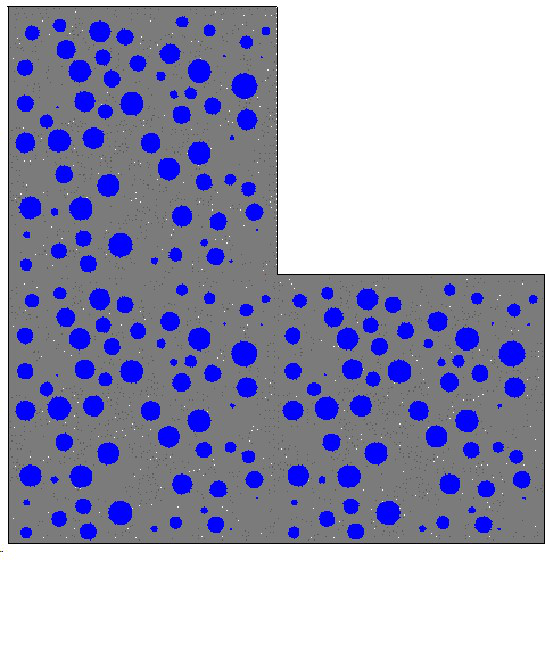}\quad\quad
    &\quad\quad\includegraphics[height=7cm]{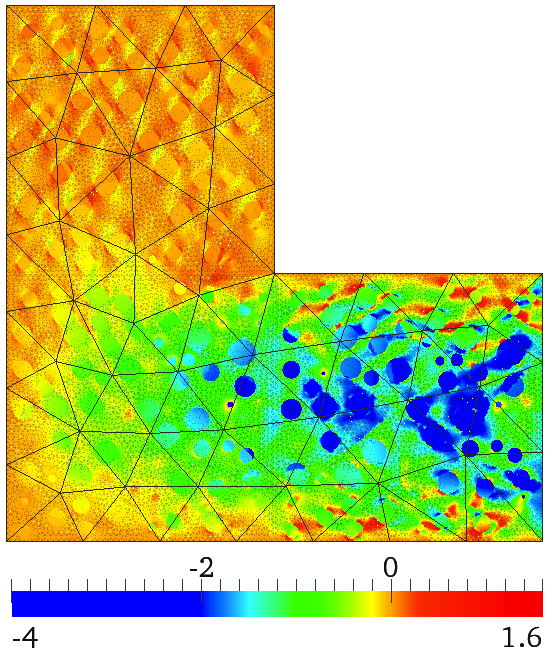}\\
    (a)&(b)\\*[2em]
    \includegraphics[height=7cm]{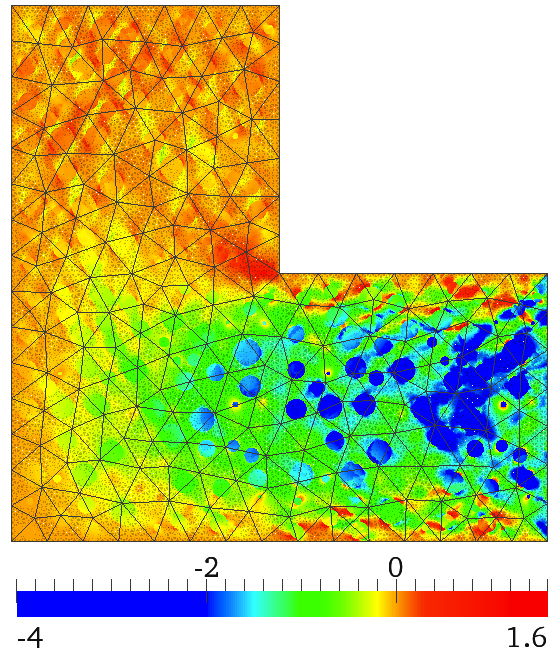}\quad\quad
    &\quad\quad\includegraphics[height=7cm]{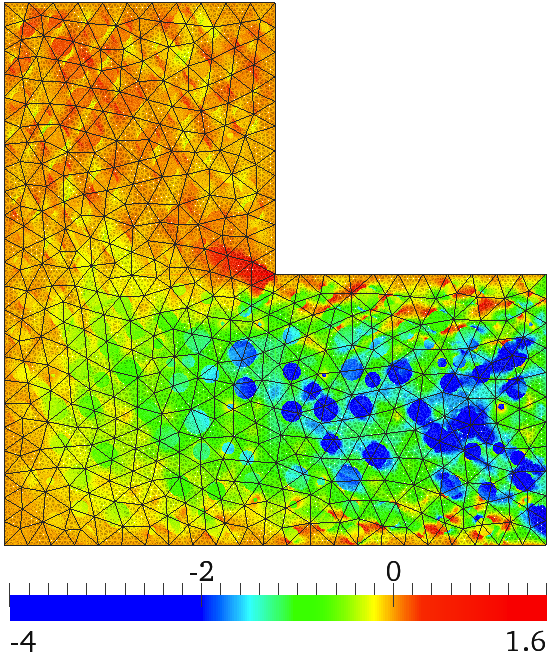}\\
    (c) &(d)
    \end{tabular}
    \caption{Solution of L-shaped specimen on $x$-$y$ mid-plane. a) microstructure. Plots of $\stress_{xy}$ resulting from b) coarse mesh, c) medium mesh and d) fine mesh.
    \label{fig:l_shape2} }
\end{figure}

The solution for three different mesh densities (\Fref{fig:l_shape1}c, d \& e)
are compared in~\Fref{fig:l_shape2}. These results are plotted on the $x$-$y$
mid-plane. \Fref{fig:l_shape2}(a--d) shows a plot of the $\stress_{xy}$ stress
component for the three different meshes. These results show that the complex
stress distribution resulting from the heterogeneities can be captured and that
the solution is converging with mesh refinement. Further local mesh refinement
near corners and stress concentrations is possible, although this was not
undertaken in this case. 
\rev{As an estimate of the computational overhead of the enhanced formulation for this particular problem on 16 processors, we note that the total solution for $30,007$ C-HTS elements was $597$s, whereas the problem on the identical mesh of HTS elements for the equivalent homogeneous problem consumed $1.6$s  of computer time. This represents a large increase of computational time in comparison to the homogenous problem. However, it should be noted that not all aspects of the solution procedure have been parallelised.}{}
It should also be noted that it was not possible to obtain
the reference solution for this problem \rev{by means of conventional FEA}{} due to the excessive number of degrees of freedom associated with a mesh required to resolve all of the heterogeneities. \rev{The mesh generation itself was not possible using our currently available software and hardware facilities. Furthermore, comparison with a PUM based solution, was also not undertaken due to the complexity of the problem.}{}
%

\section{Conclusions}
%
A new micromechanics-enhanced finite element formulation has been presented for modelling the influence of a large number of heterogeneities in composite materials in a computationally efficient manner. The strategy exploits closed form solutions derived by Eshelby for ellipsoidal inclusions in order to determine the mechanical perturbation fields as a result of the underlying heterogeneities. Approximation functions for these perturbation fields are then incorporated into a finite element formulation to augment those of the macroscopic fields. A significant feature of this approach is that the finite element mesh does not explicitly resolve the heterogeneities, \rev{although the resulting solution still explicitly accounts for their presence. In contrast with traditional homogenization approaches, this method does not rely on separation of scale and does not suffer from loss of information due to averaging or localization.}{}

The proposed technique has been implemented into a hybrid-Trefftz stress (HTS) element formulation and it has been shown that the resulting enhanced elements (C-HTS) require significantly fewer degrees of freedom to capture the detailed mechanical response compared to standard finite elements. The paper also outlines how the proposed micromechanics approach could be used within a Partition-of-Unity (PoU) formulation, although we conclude that this does not fully exploit the advantages of PoU methods and that the proposed hybrid-Trefftz formulation is most appropriate.

A \emph{self-compatibility} algorithm is used to determine the mutual interactions between inclusions, assuming that the eigenstrain fields are uniform within the domain of each inclusion. It was found that even for topologies exhibiting extremely small distances between the inclusions, this assumption is sufficient. Furthermore, it has been shown that boundary effects, that are not accounted for by the classical micromechanical solution due to the assumption of an infinite medium, can be captured through local mesh refinement.

We have implemented this formulation into our FE code that is optimized for parallel computing. Additional parallelization of the micromechanical aspects of the formulation needs to be investigated for increased efficiency. Further research is required in order to incorporate other improvements such as nonuniform eigenstrains~\cite{mura1987micromechanics}, debonding effects~\cite{ju2008micromechanical} \rev{and inclusions of arbitrary shape by evaluating the perturbation functions 
numerically~\cite{maz2007approximate,Novak:2008:CESS}}{}.

\paragraph{Acknowledgements}
Funding by the Glasgow Research Partnership in Engineering (GRPE) under project
``Multi-scale modelling of fibre reinforced composites'' is gratefully
acknowledged. This research was partially supported by the Czech Science
Foundation through project GA\v{C}R 103/09/P490 and by the Ministry of
Education, Youth and Sports of the Czech Republic trough project MSM
684077003. All the analyses were performed by means of \code{YAFFEMS} FE
code. For more details we refer to the code's homepage at
\http{code.google.com/p/yaffems}.

\appendix

\section{Detailed solution of perturbation displacements}\label{appendix:B}
The displacement perturbation field
in an infinite homogeneous material due to a uniform eigenstrain
$\strain^\tau_{ij}$ applied to an ellipsoidal region $\Omega$
is provided by the following integral
equation~\cite[Eq.~(11.30)]{mura1987micromechanics}
\begin{equation}
u^*_i = \frac{1}{8\pi (1-\nu)}\left[\Psi_{jk,jki} - 2\nu \Phi_{kk,i} -
  4(1-\nu)\Phi_{ik,k}\right]
  \label{eq:fundamental_solution_1}
\end{equation}
where $\nu$ denotes the Poisson's ratio and the elliptic potentials
$\Psi_{ij}$ and $\Phi_{ij}$ are defined
as~\cite[Eq.~(11.32)]{mura1987micromechanics}
\begin{equation}
  \begin{array}{ccc}
    \displaystyle
    \Psi_{ij} =
    \strain^\tau_{ij}\int_{\IncDomain}\abs{\vek{x}-\vek{x}'}\de{\vek{x}'}
    = \strain^\tau_{ij} \psi,
    &\mathrm{and}
    &
    \displaystyle
    \Phi_{ij} =
    \strain^\tau_{ij}\int_{\IncDomain}\frac{1}{\abs{\vek{x}-\vek{x}'}}\de{\vek{x}'}
    = \strain^\tau_{ij} \phi
  \end{array}
  \label{eq:volu_integrals}
\end{equation}
The integrals $\phi$ and $\psi$ in \Eref{eq:volu_integrals} are
the harmonic and bi-harmonic potentials respectively. Note that in
\Eref{eq:fundamental_solution_1} and thereafter, standard index notation is
employed, together with the generalised summation convection due to
\aname{Mura}~\cite{mura1987micromechanics}. Thus, a repeated
index is summed according to the \aname{Einstein} summation rule
(e.g. $a_i b_{ij} = \sum_{i=1}^3 a_i b_{ij}$), whereas a non-repeated upper-case
index equals to the lower-case equivalent (e.g. $a_i b_i c_{Ij} =
\sum_{i=1}^3 a_i b_i c_{ij}$). The symbol $a_{jk,i}$ denotes the partial
derivative of $a_{jk}$ with respect to the coordinate $x_i$.

By combining \Eref{eq:fundamental_solution_1} and \Eref{eq:volu_integrals},
we obtain
\begin{equation}
u^*_i = \frac{1}{8\pi (1-\nu)}\left[\strain^\tau_{jk}\psi_{,jki} - 2\nu
  \delta_{jk}\strain^\tau_{jk}\phi_{,i} - 4(1-\nu)\delta_{ij}\strain^\tau_{jk}\phi_{,k}\right]
  \label{eq:fundamental_solution_3}
\end{equation}
Similarly to \aname{Eshelby}'s
approach~\cite{eshelby1957determination}, the displacement perturbations are expressed in compact
form:
\begin{eqnarray}\label{eq:fundamental_solution_compact_form}
u^*_i = L_{ijk}^{\strain}\strain^\tau_{jk},
&&
L_{ijk}^{\strain} = \frac{1}{8\pi (1-\nu)}\left[\psi_{,jki} - 2\nu
  \delta_{jk}\phi_{,i} - 4(1-\nu)\delta_{ij}\phi_{,k}\right]
\end{eqnarray}
where the third-order operator $L_{ijk}$ maps a transformation
eigenstrain $\strain^\tau_{jk}$ to the displacement perturbation field $u^*_i$.  It is
therefore analogous to the well-known \aname{Eshelby}
tensor~\cite{eshelby1957determination}, which relates a
transformation eigenstrain to the strain perturbation field.

%

The operator $L_{ijk}$ can be conveniently expressed in terms of
the Ferrers-Dyson elliptic integrals,
e.g.~\cite[Eq.~(11.36)]{mura1987micromechanics}
\begin{eqnarray}
I (\lambda) & = &  2\pi a_1a_2a_3 \int_\lambda^\infty \frac{\de{s}}{\Delta (s)},\nonumber\\
I_{i}( \lambda ) & = &  2\pi a_1a_2a_3 \int_\lambda^\infty \frac{\de{s}}{(a_i^2 + s)\Delta (s)},\nonumber\\
I_{ij}( \lambda ) & = &  2\pi a_1a_2a_3 \int_\lambda^\infty \frac{\de{s}}{(a_i^2 + s)(a_j^2 + s)\Delta (s)}
\label{eq:ferer-dyson-ints}
\end{eqnarray}
where $a_i$ stands for the $i$-th semi-axis of ellipsoid $\Omega$ and
$\Delta (s)$ is obtained from
\begin{equation}
\Delta^2 (s)
=
\prod_{i=1}^3(a_i + s)^2
  \label{eq:delta-s}
\end{equation}
The variable $\lambda$ is the largest positive root of
equation~\cite[Eq.~(11.37)]{mura1987micromechanics}
\begin{equation}
\frac{x_ix_i}{(a_I + \lambda)^2} = 1
\label{eq:lambda}
\end{equation}
Notice that $\lambda$ is generally position dependent and non-zero for the
points $x_i$ placed outside the inclusion domain $\Omega$, hence called the
exterior points. Contrary, $\lambda = 0$ for interior points.

All integrals in~\eqref{eq:ferer-dyson-ints} admit a closed-form expression in
terms of the \aname{Legendre-Jacobi} integrals of the first and second kind,
defined as a fuction of an auxiliary angle
$\theta$,~\cite[Eq.~(12.17)]{mura1987micromechanics}. It is worth noting that
its definition via~\cite[Eq~(11.18)]{mura1987micromechanics}
\begin{equation}
  \theta = \sin^{-1}\sqrt{1-\frac{a_3^2}{a_1^2}}
  \label{eq:Mura's_bug}
\end{equation}
is valid for \myemph{interior points} only and not \myemph{everywhere} as
stated in~\cite{mura1987micromechanics}. Thus, it needs to be replaced with a
general formula:
\begin{equation}
  \theta = \sin^{-1}\sqrt{\frac{a_1^2 - a_3^2}{a_1^2 + \lambda}}
  \label{eq:Theta_eshelby_rahman}
\end{equation}
available, e.g. in~\cite{eshelby1959elastic,rahman2001newtonian}. Moreover, the
following identity~\cite[Eq.~(11.40.4)]{mura1987micromechanics}
\begin{equation}
\left[ x_nx_nI_{i\dots jN}(\lambda) \right]_{,p} = 2 x_p I_{i\dots jP} + I_{i\dots j,p}(\lambda)
\label{eq:mura_11.40.3a}
\end{equation}
will be repeatedly proved useful in the sequel.
%
%
%
%
%

It follows from~\Eref{eq:fundamental_solution_compact_form} that to express
operator $L^\varepsilon_{ijk}$, we need to evaluate the first-order
derivatives of the potential $\phi$ and the third-order derivatives of $\psi$.
To this end, we start with
expressions~\cite[Eq.~(11.38)]{mura1987micromechanics}
\begin{equation}\label{eq:phi_integral_using_ellp_int}
\phi(\lambda)
=
\half
\left[
  I(\lambda)
  -
  x_n x_n I_N(\lambda)
\right]
\end{equation}
and
\begin{equation}\label{eq:psi_integral_derivative_using_ellp_int}
\psi_{,i}(\lambda)
=
\half
x_i
\Bigl\{
  2 \phi(\lambda)
  -
  a^2_I
    \left[
      I_I(\lambda)
      -
      x_n x_n I_{IN}(\lambda)
    \right]
\Bigr\}
=
\half x_i Q(\lambda)
\end{equation}
Employing \Eref{eq:mura_11.40.3a}, the first derivative of $\phi$ becomes
\begin{equation}\label{eq:phi_integral_derivative_using_ellp_int}
\phi_{,i}( \lambda )
=
\half
\left\{
  I_{,i}(\lambda)
  -
  \left[
    x_n x_n I_N(\lambda)
  \right]_{,i}
\right\}
=
\half
\left[
  I_{,i}(\lambda)
  -
  2 x_i I_I(\lambda)
  -
  I_{,i}(\lambda)
\right]
=
- x_iI_I(\lambda)
\end{equation}
The third derivative of potential $\psi$ is expressed
from~\Eref{eq:psi_integral_derivative_using_ellp_int} as
\begin{equation}\label{eq:psi_integral_3rd_derivative_using_ellp_int}
\psi_{,ijk}(\lambda)
=
\half\left[
  \delta_{ij}Q_{,k}(\lambda)
  +
  \delta_{ik}Q_{,j}(\lambda)
  +
  x_i Q_{,jk}(\lambda)
\right]
\end{equation}
With the help of Eqs.~\eqref{eq:phi_integral_derivative_using_ellp_int}
and~\eqref{eq:mura_11.40.3a}, the term $Q_{,j}$ can be evaluated from
\begin{equation}\label{eq:Q_j}
Q_{,j}(\lambda)
=
2 \phi_{,j}(\lambda)
-
a^2_I
\left[
  I_I(\lambda)
  -
  x_n x_n I_{IN}(\lambda)
\right]_{,j}
=
2 x_j
\left[
  a^2_I I_{IJ}(\lambda)
  -
  I_J(\lambda)
\right]
\end{equation}
This provides the second derivatives of $Q$ in the form
\begin{equation}\label{eq:Q_jk}
  Q_{,jk}(\lambda)
  =
  2\Bigl\{
  \delta_{jk}\left[a^2_I  I_{IJ}(\lambda) -
    I_J(\lambda)\right] + x_j\left[a^2_I I_{IJ,k}(\lambda) -
    I_{J,k}(\lambda)\right]\Bigr\}
\end{equation}
After utilising the derivatives of $Q(\lambda)$ and re-ordering the indices,
\Eref{eq:psi_integral_3rd_derivative_using_ellp_int} becomes
\begin{align}\label{eq:psi_integral_3rd_derivative_using_ellp_int_components_reordered}
\psi_{jki}(\lambda)
& =
x_i\delta_{jk}
\left[
  a^2_J I_{JI}(\lambda)
  -
  I_I(\lambda)
\right]
+
x_k\delta_{ji}
\left[
  a^2_J I_{JK}(\lambda)
  -
  I_K(\lambda)
\right]
\nonumber\\
&
+ x_j\delta_{ki}
\left[
  a^2_J I_{JK}(\lambda)
  -
  I_K(\lambda)
\right]
+
x_j x_k
\left[
  a^2_J I_{JK,i}(\lambda)
  -
  I_{K,i}(\lambda)
\right]
\end{align}
with $I_{IJ,k}$ and $I_{I,j}$ provided by~\cite[Eqs. (11.40.1, 11.40)]{mura1987micromechanics}
\begin{eqnarray}\label{eq:Q_jk_2}
  I_{i\dots jk,p}(\lambda) &=& \frac{-2\pi a_1 a_2 a_3}{(a_i^2 + \lambda)\dots(a_j^2 + \lambda)(a_k^2 + \lambda)\Delta(\lambda)}\lambda_{,p},\nonumber\\
\lambda_{,p} &=& \frac{2x_p}{a_P^2 + \lambda}\frac{(a_I^2 + \lambda)^2}{x_ix_i}
\end{eqnarray}

Now we are in a position to evaluate $L_{ijk}^\strain$ in terms of the
Ferrers-Dyson integrals. Introducing
Eqs.~\eqref{eq:phi_integral_derivative_using_ellp_int}
and~\eqref{eq:psi_integral_3rd_derivative_using_ellp_int_components_reordered}
into \Eref{eq:fundamental_solution_compact_form} gives

\begin{align}\label{eq:L_ijk_full_ext_points}
\left[
  8\pi (1-\nu)
\right]
L_{ijk}^\strain
& =
x_i\delta_{jk}
\left[
  a^2_J I_{JI}(\lambda)
  -
  I_I(\lambda)
\right]
+
\left(
  x_k\delta_{ji} + x_j\delta_{ki}
\right)
\left[
  a^2_J I_{JK}(\lambda)
  -
  I_K(\lambda)
\right] \nonumber \\
& +
x_j x_k
\left[
  a^2_J I_{JK,i}(\lambda)
  -
  I_{K,i}(\lambda)
\right]
+
2\nu\delta_{jk}x_iI_I(\lambda) + 4(1 - \nu)\delta_{ij}x_kI_K(\lambda)\big\}
\end{align}
Since $\lambda=0$ for the points inside the inclusion, recall \Eref{eq:lambda}, all
derivatives of $I_{i\dots j}$ vanish as well. Therefore,
Eq.~\eqref{eq:L_ijk_full_ext_points} yields
\begin{align}\label{eq:L_ijk_full_int_points}
\left[
  8\pi (1-\nu)
\right]
L^{\strain,\mathrm{int}}
& =
x_i\delta_{jk}
\left[
  a^2_J I_{JI}(\lambda)
  -
  I_I(\lambda)
\right]
+
\left(
  x_k\delta_{ji}
  +
  x_j\delta_{ki}
\right)
\left[
  a^2_J I_{JK}(\lambda)
  -
  I_K(\lambda)
\right]
\nonumber \\
& +
\nu \delta_{jk} x_i I_I(\lambda)
+
4(1 - \nu)\delta_{ij}x_kI_K(\lambda)
\end{align}
and \Eref{eq:fundamental_solution_compact_form} receives its final form
\begin{equation}\label{eq:L_ijk_full_ext_points_L^int}
L_{ijk}^\strain
=
L^{\strain,\mathrm{int}}_{ijk}
+
\frac{1}{8\pi (1-\nu)}
x_j x_k
\left[
  a^2_J I_{JK,i}(\lambda)
  -
  I_{K,i}(\lambda)
\right]
\end{equation}
For implementation purposes, it is worth noting that
Eq.~\eqref{eq:fundamental_solution_compact_form}
admits the Voigt representation:
\begin{equation}
  \left\{\begin{array}{c}
  u^*_1\\
  u^*_2\\
  u^*_3
  \end{array}\right\}
  =
  \left[\begin{array}{cccccc}
      L_{111} &L_{122} &L_{133} &L_{112} &L_{123} &L_{113}\\
      L_{211} &L_{222} &L_{233} &L_{212} &L_{223} &L_{213}\\
      L_{311} &L_{322} &L_{333} &L_{312} &L_{323} &L_{313}
    \end{array}\right]
  \left\{\begin{array}{r}
  \strain^\tau_{11}\\
  \strain^\tau_{22}\\
  \strain^\tau_{33}\\
  2\strain^\tau_{12}\\
  2\strain^\tau_{23}\\
  2\strain^\tau_{13}
  \end{array}\right\}
  \label{eq:u_i=L_ijk:e_jk}
\end{equation}
%


\end{document}